\documentclass[a4paper,12pt]{article}
\usepackage{amsfonts,amsthm,amsmath,amssymb,graphicx,hyperref}
\usepackage[vcentermath]{youngtab}

\def\Tr{{\rm Tr\, }}

\newcommand{\be}{\begin{equation}}
\newcommand{\bea}{\begin{eqnarray}}
\newcommand{\ee}{\end{equation}}
\newcommand{\eea}{\end{eqnarray}}

\begin{document} 

\makeatletter
\@addtoreset{equation}{section}
\makeatother
\renewcommand{\theequation}{\thesection.\arabic{equation}}
\vspace{1.8truecm}

{\LARGE{ \centerline{\bf Exciting LLM Geometries}         }}  

\vskip.5cm 

\thispagestyle{empty} 
\centerline{ {\large\bf Robert de Mello Koch$^{a,b,}$\footnote{{\tt robert@neo.phys.wits.ac.za}},
Jia-Hui Huang$^{a,}$\footnote{{\tt huangjh@m.scnu.edu.cn}} }}

\centerline{ {\large\bf and Laila Tribelhorn${}^{b,}$\footnote{ {\tt laila.tribelhorn@gmail.com}}}}

\vspace{.4cm}
\centerline{{\it ${}^a$ School of Physics and Telecommunication Engineering},}
\centerline{{ \it South China Normal University, Guangzhou 510006, China}}

\vspace{.4cm}
\centerline{{\it ${}^b$ National Institute for Theoretical Physics,}}
\centerline{{\it School of Physics and Mandelstam Institute for Theoretical Physics,}}
\centerline{{\it University of the Witwatersrand, Wits, 2050, } }
\centerline{{\it South Africa } }

\vspace{1truecm}

\thispagestyle{empty}

\centerline{\bf ABSTRACT}

\vskip.2cm 

We study excitations of LLM geometries.
These geometries arise from the backreaction of a condensate of giant gravitons.
Excitations of the condensed branes are open strings, which give rise to an emergent Yang-Mills theory at low energy. 
We study the dynamics of the planar limit of these emergent gauge theories, accumulating evidence that they are planar ${\cal N}=4$ super Yang-Mills.
There are three observations supporting this conclusion: (i) we argue for an isomorphism between the planar
Hilbert space of the original ${\cal N}=4$ super Yang-Mills and the planar Hilbert space of the emergent gauge theory,
(ii) we argue that the OPE coefficients of the planar limit of the emergent gauge theory vanish and (iii) we argue that the
planar spectrum of anomalous dimensions of the emergent gauge theory is that of planar ${\cal N}=4$ super Yang-Mills.
Despite the fact that the planar limit of the emergent gauge theory is planar ${\cal N}=4$ super Yang-Mills, we explain why the emergent gauge theory is not ${\cal N}=4$ super Yang-Mills theory.

\setcounter{page}{0}
\setcounter{tocdepth}{2}
\newpage
\tableofcontents
\setcounter{footnote}{0}
\linespread{1.1}
\parskip 4pt

{}~
{}~

\section{Introduction}

The map between the planar limit of ${\cal N}=4$ super Yang-Mills theory and an integrable spin chain\cite{Minahan:2002ve}
has been a surprisingly rich idea.
Single trace operators in the conformal field theory (CFT) are identified with states of the spin chain,  and the dilatation
operator of the CFT with the Hamiltonian of the spin chain.
This allows the exact computation of anomalous dimensions and hence precision tests\cite{Gromov:2013pga,Beisert:2010jr} 
of the duality with string theory on AdS$_5\times$S$^5$\cite{Maldacena:1997re,Gubser:1998bc,Witten:1998qj}.
Excitations of the closed string are identified as magnons.
The magnons are visible in the dual string theory description\cite{Berenstein:2005jq,Hofman:2006xt}.
After projecting the closed string solution to a plane (the so called bubbling plane \cite{Lin:2004nb}) and using coordinates 
suited to 1/2 BPS supergravity geometries, the string worldsheet traces out a polygon\cite{Hofman:2006xt}.
The sides of the polygon are the magnons. 
Geometrical properties of these sides (their length and orientation) determine the conserved 
charges (momentum and energy) labeling the magnon.
The S-matrix for magnon scattering is determined up to a single overall phase simply by kinematics\cite{Beisert:2005tm}.
Integrability then fixes this phase.
The S-matrix computed in string theory is in exact agreement with the S-matrix computed in the CFT.

How much, if anything, of this story survives for string excitations of new geometries?
The geometries that we have in mind are the LLM geometries\cite{Lin:2004nb}.
An LLM geometry is dual to an operator with a dimension that grows as $N^2$ in the large $N$ limit.
Consequently, correlators of operators with dimensions of order $N^2$ encode the physics of excitations of these geometries.
For operators with such a large dimension the planar approximation is not justified\cite{Balasubramanian:2001nh}.
Consequently, mixing between different trace structures is not suppressed.
The identification between single trace operators in the CFT and spin chain states is spoiled and it seems that the link 
to an integrable spin chain is lost. 
In this introduction we will give some physical arguments which suggest that, at least for a subset of excitations, this is not the case.
The rest of the paper then carries out detailed CFT computations that confirm the details of this physical picture.

The LLM geometries are dual to a 1/2 BPS sector of the CFT.
This 1/2 BPS sector contains all gauge invariant operators built from a single complex matrix $Z$.
Since we study single matrix dynamics, there is a simple free fermion description, obtained by working in terms of the
eigenvalues of $Z$\cite{Brezin:1977sv,Corley:2001zk}.
There is also a closely related description which employs Schur polynomials in $Z$\cite{Corley:2001zk,Berenstein:2004kk}.
We mainly use this second description as we know how to generalize it when including more 
matrices\cite{Bhattacharyya:2008rb,Bhattacharyya:2008xy}.
This is needed when studying small fluctuations of the LLM geometries.
A Schur polynomial dual to an LLM geometry is labeled by a Young diagram with order $N^2$ boxes\cite{Lin:2004nb}.
An operator dual to a smooth supergravity geometry has a Young diagram with $O(1)$ corners and the distance 
between any two adjacent corners (that is, the number of rows or columns ending on the side between 
the two corners) is order $N$.
The string theory understanding of this geometry is that it is the state obtained from back reaction of condensed giant
gravitons\cite{McGreevy:2000cw,Hashimoto:2000zp,Grisaru:2000zn}.
The translation between the CFT and string theory descriptions is direct: we read the rows of the Young diagram as 
dual giant gravitons or the columns as giant gravitons\cite{Corley:2001zk}.

To excite the geometry in the CFT description, add boxes at a particular corner of the Young diagram describing
the LLM geometry\cite{Koch:2008ah,Koch:2016jnm,deMelloKoch:2018tlb}.
In string theory we understand this as exciting the giants that condensed to produce the geometry.
The description of worldvolume excitations of these D3 brane giant gravitons is in terms of some open string field theory 
whose low energy limit gives rise to a new emergent Yang-Mills theory\cite{Balasubramanian:2002sa,Balasubramanian:2004nb}.
Relative to the original Yang-Mills theory we started with, the space of the giant's worldvolume is an emergent space. The new emergent Yang-Mills theory may itself have a holographic description so we might have new holographic dualities in this large charge limit\cite{Balasubramanian:2002sa}.

The intuitive picture sketched above suggests that excitations arising from any particular corner give rise to a distinct super Yang-Mills theory.
We will study the planar limit of these emergent gauge theories, to provide detailed support for this intuition.
To restrict to the planar limit consider excitations with a bare dimension of at most $O(\sqrt{N})$, i.e. add
at most $O(\sqrt{N})$ boxes to any given corner.
Concretely we will demonstrate three things

\begin{itemize}

\item[1.] An isomorphism between the planar Hilbert space of the original ${\cal N}=4$ super Yang-Mills theory and
the planar Hilbert space of the emergent gauge theory arising at a corner. 
When restricted to the 1/2 BPS sector, these Hilbert spaces are in fact a generalization of the code subspaces constructed 
by \cite{Berenstein:2017rrx} (see also \cite{Berenstein:2017abm,Lin:2017dnz,Simon:2018laf}).

\item[2.] Three point functions of operators in the planar emergent gauge theory vanish. 
We demonstrate this in the free field theory. In the planar limit of matrix models
the vanishing follows because to mix three single traces we have to break some index loops which costs (at least) a factor of
$N$. This is a general conclusion true for both free and interacting matrix models. Consequently we conjecture that our free field theory result holds after interactions are turned on. Since operator product expansion (OPE) coefficients can be read from the three point functions, this implies the OPE coefficients of the planar emergent gauge theory vanish.

\item[3.] The correct spectrum of planar anomalous dimensions of the emergent gauge theory. 
We know the planar spectrum of anomalous dimensions of ${\cal N}=4$ super Yang-Mills theory. 
We find the same spectrum for the emergent gauge theory. 
This demonstrates integrability for the emergent gauge theories.

\end{itemize}

Notice that since any CFT is determined by its spectrum of anomalous dimensions and OPE coefficients, and that
in the strict planar limit all OPE coefficients vanish, this demonstrates that the planar limit of the emergent gauge theories are
planar ${\cal N}=4$ super Yang-Mills theory.
We will see that although these different emergent gauge theories all share the same coupling constant (which is expected
since this coupling is equal to the string coupling constant of the original string theory on AdS$_5\times$S$^5$), they
generically have distinct gauge groups $U(N_{\rm eff})$.
The rank of the gauge group $N_{\rm eff}$ receives contributions both from the flux of the original $N$ D3 branes 
that gives rise to the ${\cal N}=4$ super Yang-Mills theory we start with and from the giants which have condensed.
By considering a large charge state, its possible to have an emergent gauge theory with gauge group that has rank larger than $N$.

What we are finding is that a subset of the excitations of large charge states of the ${\cal N}=4$ super Yang-Mills theory
are equivalent to excitations of the vacuum.
There are of course excitations that go beyond the planar limit of the emergent gauge theory.
The excitation is constructed by adding boxes to the Young diagram describing the LLM geometry.
We might add so many boxes that we reach beyond two corners of the Young diagram defining the LLM geometry.
The excitation is ``too big'' to sit on the Young diagram and in this way we can detect features of the background Young diagram.
These excitations are obtained by adding $\sim N$ boxes and hence do not belong to the planar limit of the 
emergent gauge theory - they are giant graviton like operators of the emergent theory.
There are also excitations constructed by adding order $\sqrt{N}$ boxes, with the boxes added at different 
corners\cite{Koch:2015pga,Koch:2016jnm,deMelloKoch:2018tlb}.
These (delocalized) states can be described as strings with magnon excitations that stretch between two corners. 
We will show that at large $N$ these states are decoupled from (localized) states in the planar Hilbert space of the emergent
gauge theory, so that if we start from a state in the planar Hilbert space, the large $N$ dynamics will not take us out of this space.
This is an important point to demonstrate since the coupling of the planar Hilbert space of the emergent gauge theory 
to other degrees of freedom will almost certainly ruin integrability.

\begin{center}
\begin{figure}[!h]
\centering
\includegraphics[width=0.7\textwidth]{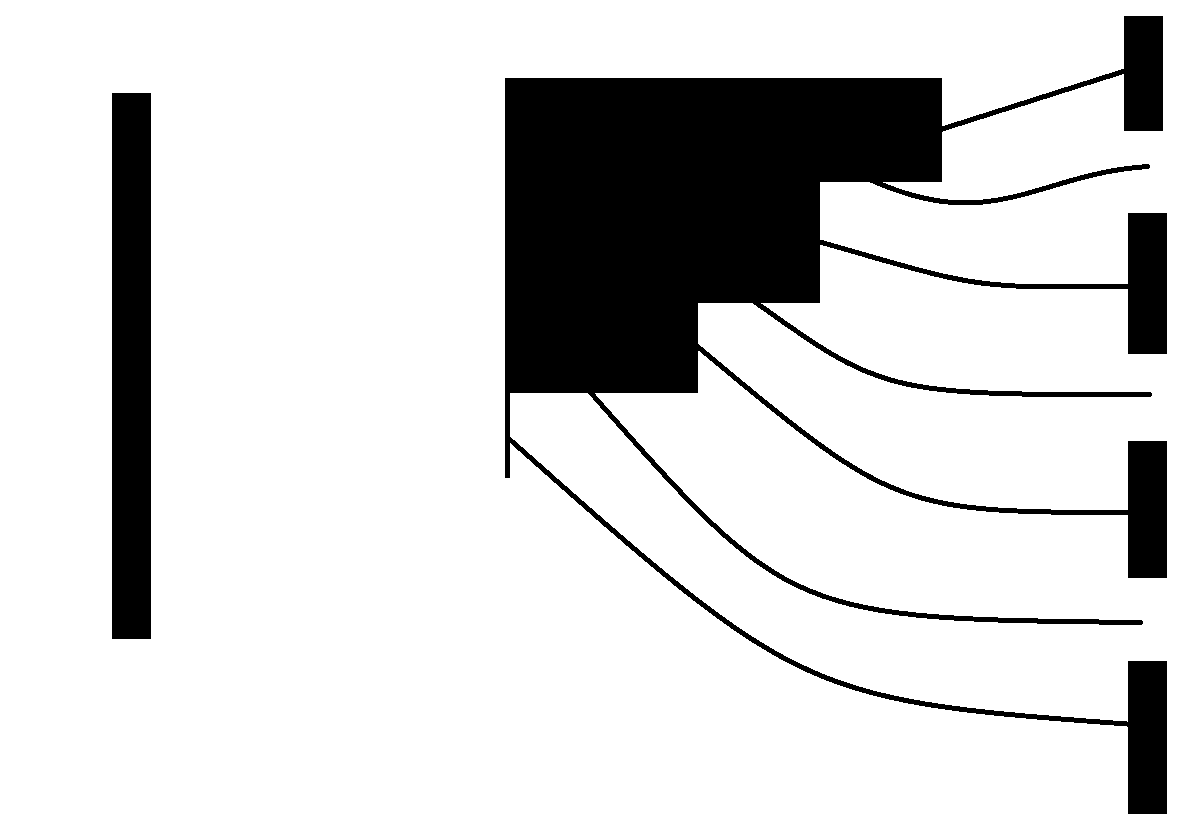}
\caption{The free fermion description of a state labeled by a Young diagram. On the left we have the Fermi sea corresponding
to the AdS$_5\times$S$^5$ geometry. The states are simply filled from the lowest to highest energy with no unoccupied
states. On the right, the Young diagram corresponding to a particular LLM geometry is shown. Each vertical edge of the 
Young diagram maps into occupied states while the horizontal edges map into unoccupied states. The number of fermions 
that were not excited at all is equal to the number of rows with no boxes. Thus, the excited state has broken the Fermi sea 
up into a series of occupied blocks.}
\label{fermipicture}
\end{figure}
\end{center}

The free fermion description of the system is a powerful description because of its simplicity.
The large charge state corresponds to exciting the fermions as illustrated in Fig \ref{fermipicture}.
The idea that a subset of the excitations of large charge states of the CFT are equivalent to excitations
of the vacuum has a natural interpretation in this free fermion language.
We are saying that exciting any edge of the blocks appearing in the excited state is equivalent to exciting the edge 
of the original Fermi sea.
The only difference between the different blocks is their extent.
By restricting to the planar limit we consider excitations that are not able to detect that the Fermi sea is not
infinite, so the extent of each block is irrelevant.

We will be using group representation theory methods to approach the problem of computing correlators of operators with a bare dimension of order $N^2$.
This approach has been developed in a series of articles\cite{Corley:2001zk},\cite{deMelloKoch:2007uu}-\cite{Kimura:2012hp},\cite{Bhattacharyya:2008rb},\cite{Bhattacharyya:2008xy},
which has developed a number of bases for the local operators of the theory.
These bases diagonalize the two point function of the free theory to all orders in $1/N$, and they mix weakly at
weak coupling\cite{deMelloKoch:2007nbd,DeComarmond:2010ie,Brown:2008rs}.
They therefore provide a very convenient tool with which to tackle the large $N$ but non-planar limit of the CFT.

The representation theory methods sum the complete set of ribbon graphs.
In this approach, operators are constructed using projection operators\footnote{These operators are actually intertwiners since they map between different copies of the representations involved. For simplicity though the reader may think of them as projectors which are more familiar.} of the symmetric group
so that the gauge invariant operators are labeled with irreducible representations of the group.
Summing the ribbon diagrams of the free theory becomes multiplying these projectors and then taking a trace.
At loop level, we evaluate the dilatation operator $D$.
Evaluating matrix elements of $D$ amounts to computing the trace of the product of commutators of elements of the
symmetric group with projection operators.
The central technical achievement is that in the end computing correlators, i.e. summing the ribbon graphs, is reduced to 
well defined (but technically involved) problems in group representation theory.
A helpful point of view in making sense of the details, which we introduce and develop in this article, entails classifying 
the various ingredients of the computation as background independent or background dependent.
By something that is background independent, we mean something that would take the same value on
any inward pointing corner of any Young diagram dual to an LLM geometry, or even in the absence of a background, i.e.
in the planar limit of the original CFT.
These are quantities that take the same value regardless of which collection of branes we excite, and this is what we
signify in the terminology ``background independent''.
A quantity that is background dependent does depend on the collection of branes we excite.
As we discuss in section \ref{freecft}, after making this distinction it is clear that the Hilbert spaces of the planar limit
of the emergent gauge theory at any corner are isomorphic to each other and to the planar Hilbert space of 
the theory in the absence of a background.

One of the original motivations for this study are the results \cite{Koch:2016jnm,deMelloKoch:2018tlb,Kim:2018gwx} which suggest the existence of new integrable subsectors of the CFT.
We want to explore (and further establish) the existence of these integrable subsectors.
As discussed above, a key issue is to understand if the integrable sectors are decoupled from the nonintegrable sectors.
It is useful to bear in mind that integrability in the planar limit also depends on a decoupling between different subspaces: 
it makes use of the fact that different trace structures don't mix.
Thanks to this decoupling, it is consistent to focus on the space of single trace operators and it is in this subspace that 
its possible to construct a bijection between operators and the states of an integrable spin chain.
The statement of this decoupling is coded into the planar correlation functions: correlators of operators with different
trace structures vanish as $N\to\infty$.
Motivated with this insight we focus on correlation functions of the large $N$ but non-planar limits to establish
the decoupling between integrable and non-integrable subsectors.
This is discussed in section 2, where we obtain a simple formula for the correlators in the planar limit of the free 
emergent gauge theory in terms of correlators of the free planar CFT without background.
Consequently the decoupling we establish is closely related to the absence of mixing between different trace structures
in the planar limit.

We extend these results to the weakly interacting CFT in section 3, giving arguments that the spectrum
of planar anomalous dimensions of the emergent gauge theory match the spectrum of planar anomalous dimensions
of the original ${\cal N}=4$ super Yang-Mills theory.
We revisit the issue of coupling between integrable and non-integrable subsectors, arriving at the conclusion that
the two are decoupled even after interactions are turned on.

In section 4 we consider the strongly coupled CFT, using the dual string description.
We explain why the excitations considered should be understood as open string excitations
localized on the world volume of giant graviton branes.
We also suggest how to describe closed string excitations of the large charge state we consider.
In section 5 we summarize and discuss a number of promising directions in which to extend this work.

\section{Free CFT}\label{freecft}

Our basic goal is to organize and study excitations of an LLM background, using the dual ${\cal N}=4$ super Yang-Mills
theory.
Any LLM geometry is specified by a boundary condition, given by coloring the bubbling plane into black and white regions\cite{Lin:2004nb}.
The LLM backgrounds we consider have boundary conditions given by concentric annuli, possibly with a central black disk.
The LLM geometry is described by a CFT operator with a bare dimension of order $N^2$.
Concretely, it is a Schur polynomial\cite{Corley:2001zk} labeled by a Young diagram with $O(N^2)$ boxes and $O(1)$ corners.
Large $N$ correlators of these operators are not captured by summing only planar diagrams, so we talk about the large 
$N$ but non-planar limit of the theory.
The excitation is described by adding $J$ boxes to the background, with $J^2\ll N$.
Consequently, we can ignore back reaction of the excitation on the LLM geometry.

The CFT operators corresponding to the background and excitation are given by restricted Schur polynomials\cite{Bhattacharyya:2008rb,Bhattacharyya:2008xy}.
Construction of these operators and their correlators becomes an exercise in group representation theory.
In section \ref{bckindep} we discuss elements of this description, placing an emphasis on if the 
quantity being considered depends on or is independent of the collection of branes being excited.
This distinction will clarify general patterns in the CFT computations that follow. 

We begin our study in the free field theory.
The Hilbert space of possible excitations can be written as a direct sum of subspaces.
There are subspaces that collect the excitations localized at the outer or inner edge of a given annulus, or at the
outer edge of the central disk.
The excitations are obtained by adding boxes to the Young diagram describing the background, at a 
specific location.
They are also localized in the dual gravitational description, at a specific radius on the bubbling plane\cite{Koch:2016jnm,deMelloKoch:2018tlb}.
Each localized Hilbert space is labeled by the edge at which it is localized.
There are also delocalized excitations, where the description of the excitation involves adding boxes at
different locations on the background Young diagram\cite{Koch:2016jnm,deMelloKoch:2018tlb}.
We will not have much to say about delocalized excitations.

The excitations belonging to the localized Hilbert spaces play a central role in our study.
These are the Hilbert spaces of the emergent gauge theories.
We give a bijection between the states belonging to the planar Hilbert space of an emergent gauge theory, and the states of the planar limit of the original CFT without background.
To show that the bijection takes on a physical meaning, we argue that correlation functions of operators that are in bijection are related in a particularly simple way, in the large $N$ limit.
This result is significant because the basic observables of any quantum field theory are its correlation functions and many
properties of the theory can be phrased as statements about correlation functions.
Thanks to the map between correlation functions, any statement about the planar limit that can be phrased in terms
of correlators, immediately becomes a statement about the planar emergent gauge theories that arise in the large $N$ but non-planar limits we consider.

\subsection{Background Dependence}\label{bckindep}

Irreducible representations of the symmetric group $S_n$ are labeled by Young diagrams with $n$ boxes.
States in the carrier space of the representation are labeled by standard tableau, in which we populate the
boxes with numbers $\{1,2,...,n\}$ such that the numbers are decreasing along the rows (from left to right)
and along the columns (from top to bottom).
A representation for $S_n$ is given by specifying the action of any element $\sigma\in S_n$ on the standard
tableau.
We will use Young's orthogonal representation.
For example, here is the action of the two cycle $\sigma =(12)\in S_4$ on a specific tableau
\bea
\text{\large{(12)}}\,\,\,\young(431,2)={1\over 3}\,\,\young(431,2)+\sqrt{1-\left({1\over 3}\right)^2}\,\,\young(432,1)
\eea
The number ${1\over 3}$ that appears in the above equation is counting the number of boxes in the shortest path
from the box labeled 1 to the box labeled 2.
The only thing that matters is the relative position of boxes 1 and 2.
Consequently, $\sigma$ has the same action on all three states shown below.
\bea
{\footnotesize \young(431,2)}\qquad
{\footnotesize \young(\,\,\,\,\,\,\,\,\,431,\,\,\,\,\,\,\,\,\,2,\,\,\,\,\,\,\,\,\,,\,\,\,\,\,\,,\,\,\,\,\,\,,\,\,\,\,\,\,,\,\,\,,\,\,\,,\,\,\,)}
\qquad
{\footnotesize \young(\,\,\,\,\,\,\,\,\,,\,\,\,\,\,\,\,\,\,,\,\,\,\,\,\,\,\,\,,\,\,\,\,\,\,,\,\,\,\,\,\,,\,\,\,\,\,\,,\,\,\,431,\,\,\,2,\,\,\,)}
\eea
For example
\bea
\text{\large{(12)}}\,\,\,
{\footnotesize \young(\,\,\,\,\,\,\,\,\,,\,\,\,\,\,\,\,\,\,,\,\,\,\,\,\,\,\,\,,\,\,\,\,\,\,,\,\,\,\,\,\,,\,\,\,\,\,\,,\,\,\,431,\,\,\,2,\,\,\,)}=
{1\over 3}
{\footnotesize \young(\,\,\,\,\,\,\,\,\,,\,\,\,\,\,\,\,\,\,,\,\,\,\,\,\,\,\,\,,\,\,\,\,\,\,,\,\,\,\,\,\,,\,\,\,\,\,\,,\,\,\,431,\,\,\,2,\,\,\,)}
+\sqrt{1-\left({1\over 3}\right)^2}
{\footnotesize \young(\,\,\,\,\,\,\,\,\,,\,\,\,\,\,\,\,\,\,,\,\,\,\,\,\,\,\,\,,\,\,\,\,\,\,,\,\,\,\,\,\,,\,\,\,\,\,\,,\,\,\,432,\,\,\,1,\,\,\,)}
\eea
This demonstrates that the action of the symmetric group on the boxes belonging to the excitation is background
independent.
In what follows we will use $R$ (or $r$) to denote the Young diagram describing the excitation and $+R$ (or $+r$)
to denote the Young diagram after it has been placed at an inward pointing corner of the Young diagram for the LLM
geometry.

When our excitation has more than one type of field, the gauge invariant operator is constructed by restricting to the 
subgroup that permutes fields of a specific type.
For example, if we have $n$ $Z$ fields and $m$ $Y$ fields, we would start with an irreducible representation
$R\vdash n+m$ of $S_{n+m}$ and restrict to some representation $(r,s),$ $r\vdash n,$ $s\vdash m,$ of the 
$S_n\times S_m$ subgroup.
Upon restricting $(r,s)$ may appear more than once, so we need a multiplicity label $\alpha$ to distinguish the 
different copies.
Since we use only the action of the symmetric group to perform the restrictions, the multiplicity labels are also background
independent.
To diagonalize the one loop dilatation operator \cite{Koch:2011hb,deMelloKoch:2012ck} traded the multiplicity labels for directed graphs recording how open strings are connected between giant gravitons.
These graphs summarize basic physics coming from the Gauss Law on the brane worldvolume that is true for any collection of compact branes.
This is why the multiplicity labels are background independent.

There is a potential fly in the ointment that deserves discussion.
In the absence of the background, $R$ is used to put the $Z$s and $Y$s together while $r$ is used to organize
the $Z$s and $s$ the $Y$s. 
In the presence of the background, constructed using $Z$s, we must replace $R\to +R$ and $r\to +r$, while $s$ is
unchanged. 
The first $m$ boxes labeled in the standard tableau made by filling $R$ are $Y$ fields, and are among the impurity
boxes added to the background Young diagram.
The remaining boxes are then labeled in all possible ways to give the states of the subspace. Imagine that $n=m=2$.
Two possible labeling are as follows
\bea
{\footnotesize \young(\,\,\,\,\,\,\,\,\,,\,\,\,\,\,\,\,\,\,,\,\,\,\,\,\,\,\,\,,\,\,\,\,\,\,,\,\,\,\,\,\,,\,\,\,\,\,\,,\,\,\,431,\,\,\,2,\,\,\,)}
\qquad\qquad
{\footnotesize \young(\,\,\,\,\,\,\,\,\,,\,\,\,\,\,\,\,\,\,,\,\,\,\,\,\,\,\,4,\,\,\,\,\,\,,\,\,\,\,\,\,,\,\,\,\,\,\,,\,\,\,\,31,\,\,\,2,\,\,\,)}
\eea
On the left we have the usual action of the symmetric group on the added boxes. 
For the state on the right, we find a different answer. 
At large $N$, when the number of boxes in the shortest path linking distant box 4 to any local labeled box (where the excitation was added) is of order $N$,
any permutation swapping box 4 with another box, will just swap the two labels.
This is orthogonal to the state before the swap.
We will always land up taking a trace over group elements of the subgroup that permutes excitation boxes.
For the traces we need only states on the left contribute.
As a consequence, although the action of the symmetric group on impurities is not background independent traces 
over these elements are\footnote{This is of course up to a factor which is determined by the dimension of the irreducible representation of the background. This factor is from summing over all the possible standard tableau obtained by filling boxes associated to the background.}. 
Notice that the problem of resolving multiplicities is phrased entirely in terms of the subgroup acting on $Y$ fields i.e. we can set the problem up so that the multiplicities are associated to representation $s$.
For this reason the above potential spanner in the works doesn't threaten our conclusion that multiplicity labels
are background independent.

The operators which generalize the Schur polynomials when more than one type of field is present are called restricted
Schur polynomials\cite{deMelloKoch:2007uu,Bhattacharyya:2008rb}.
The Schur polynomial is constructed using characters of the symmetric group.
The restricted Schur polynomial is constructed using a restricted character 
$\chi_{R,(r,s),\alpha\beta}(\sigma)$ \cite{deMelloKoch:2007uu}.
Recall that the character $\chi_R(\sigma)$ is given as a trace over the matrix $\Gamma_R(\sigma)$ representing
$\sigma$ in irreducible representation $R$.
For the restricted character we restrict the trace to the subspace carrying the representation of the subgroup $(r,s)$.
Because there are different copies of $(r,s)$ in the game, there are many ways to do this.
The restricted character $\chi_{R,(r,s)\alpha\beta}(\sigma )$ is given by summing the row index of 
$\Gamma_R(\sigma)$ over the $\alpha$ copy of $(r,s)$ and the column label over the $\beta$ copy 
of $(r,s)$.
This can be accomplished by making use of an intertwining map $P_{R,(r,s)\alpha\beta}$ which maps from the
$\alpha$ copy of $(r,s)$ to the $\beta$ copy of $(r,s)$.
This map can be constructed using only elements of the symmetric group that act on the impurities.
In terms of $P_{R,(r,s)\alpha\beta}$ we have
\bea
   \chi_{R,(r,s)\alpha\beta}(\sigma ) ={\rm Tr}\left(P_{R,(r,s)\alpha\beta}\,\Gamma_R(\sigma)\right)\label{rcnb}
\eea
In the presence of the background this becomes
\bea
   \chi_{+R,(+r,s)\alpha\beta}(\sigma ) ={\rm Tr}\left(P_{+R,(+r,s)\alpha\beta}\,\Gamma_{+R}(\sigma)\right)
\eea
where $\sigma$ is the same permutation as in (\ref{rcnb}).
It is clear that the restricted character is background independent, up to the remark of footnote 2.

The operators of the planar limit are dual to strings and gravitons in the AdS$_5\times$S$^5$ geometry.
Since the restricted Schur polynomials provide a basis, any such operator can be expressed as a linear combination 
of restricted Schurs.
For simplicity we will discuss operators constructed from two complex matrices $Z$ and $Y$, but it will be clear
that our conclusions generalize for an arbitrary local operator.
The definition of the restricted Schur polynomial is\cite{Bhattacharyya:2008rb}
\begin{equation}
\chi_{R,(r,s)\alpha\beta}(Z,Y)={1\over n!m!}\sum_{\sigma\in S_{n+m}}\chi_{R,(r,s),\alpha\beta}(\sigma )
Y^{i_1}_{i_{\sigma (1)}}\cdots Y^{i_m}_{i_{\sigma (m)}}
Z^{i_{m+1}}_{i_{\sigma (m+1)}}\cdots Z^{i_{n+m}}_{i_{\sigma (n+m)}}\, .
\label{restrictedschur}
\end{equation}
An arbitrary operator $O_A$ can be expanded in the basis of restricted Schur polynomials as follows\cite{Bhattacharyya:2008xy}
\bea
O_A = \sum_{R,r,s,\alpha,\beta} a^{(A)}_{R,(r,s),\alpha,\beta} \chi_{R,(r,s)\alpha\beta}(Z,Y,X,\cdots)
\eea
We will argue that the expansion coefficients $a^{(A)}_{R,(r,s),\alpha,\beta}$ are background independent.
Imagine that $O_A$ is the operator in the planar Hilbert space corresponding to some specific state, labeled by its 
dimension, ${\cal R}$-charge and whatever other labels we need to specify it completely.
The operator in the planar Hilbert space of the emergent gauge theory, dual to the state that shares the same labels, is given by
\bea
O_{+A} = \sum_{R,r,s,\alpha,\beta} a^{(A)}_{R,(r,s),\alpha,\beta} \chi_{+R,(+r,s)\alpha\beta}(Z,Y,X,\cdots)\label{ib}
\eea
It is in this sense that the expansion coefficients are background independent.
We will argue for (\ref{ib}) below by demonstrating that with this rule the correlation functions of the set of operators 
$\{O_{+A}\}$ are given in terms of those of $\{O_A\}$, essentially by replacing $N\to N_{\rm eff}$. The two operators should then represent the same physical state since the physical interpretation of any operator is coded into its correlation functions.

We now consider quantities that are background dependent.
The two point function of restricted Schur polynomials includes a product of the factors of the Young diagram.
A box in row $i$ and column $j$ of a Young diagram has factor $N-i+j$.
This quantity clearly depends sensitively on where you are located within the Young diagram and is not simply
a function of the relative position of two boxes.
The factors of the boxes added at different corners will depend on the corner and on the details of the shape
of the Young diagram.
We will see in what follows that all of the $N$ dependence of the correlators comes from factors, so that
moving between different corners shifts $N\to N_{\rm eff}$, which changes the rank of the emergent gauge group.
The only difference between the planar limit of the emergent gauge theories at each corner is this shift in $N$.

A second ingredient in the two point function of restricted Schur polynomials, is a ratio of the product of the
hook lengths of the Young diagram.
Assume that we have a total of $C$ outward pointing corners and further that our localized excitation
is stacked in the $i$th corner.
In the Appendix \ref{ProveIdentity} we prove the following result
\bea
{{\rm hooks}_{+R}\over {\rm hooks}_{+r}}={{\rm hooks}_{R}\over {\rm hooks}_{r}}
(\eta_B)^{|R|-|r|}\left( 1+O\left({1\over N}\right)\right)\label{HIdentity}
\eea
where $|R|$ stands for the number of boxes in the Young diagram $R$ and
\bea
   \eta_B=\prod_{j=1}^i {L(j,i)\over L(j,i)-N_j}\prod_{l=i+1}^C {L(i+1,l)\over L(c+1,l)-M_l}
\eea
\bea
  L(a,b)=\sum_{k=a}^b (N_k+M_k)
\eea
The notation in the above formulas is defined in Figure \ref{LLM}.
Formula (\ref{HIdentity}) is telling us that although ${\rm hooks}_{+R}/{\rm hooks}_{+r}$ depends
on the background this dependence is a simple multiplicative factor that is sensitive to the shape of the Young diagram
for the LLM geometry and the number of fields in the excitation that are not $Z$ fields.
Its dependence on $R$ and $r$ nicely matches ${\rm hooks}_{R}/{\rm hooks}_{r}$.
Note that (\ref{HIdentity}) is not exact - it receives ${1\over N}$ corrections.

Our discussion in this section has focused on operators constructed using only 2 fields, $Z$ and $Y$.
The generalization is straight forward.
For $k$ different species of fields (which may include additional scalars, fermions or covariant derivatives),
with $n_k$ fields of each species, we consider a subgroup $S_{n_1}\times S_{n_2}\times\cdots\times S_{n_k}$
of $S_{n_1+n_2+\cdots+n_k}$.
By including enough different species we can describe any operator in the planar limit of the CFT.
It is again clear that although the action of the symmetric group on impurities is not background independent, traces 
over these elements are and that multiplicity labels and expansion coefficients are again background independent.

\subsection{Excitations of AdS$_5\times$S$^5$}

Start in the simplest setting in which no giant graviton branes have condensed and consider excitations that are dual 
to operators with a bare dimension of order $J$ with $J^2\ll N$.
This corresponds to the planar limit of the ${\cal N}=4$ super Yang-Mills.
In this limit there are important simplifications.
First, different trace structures don't mix\footnote{For a careful study of this point see \cite{Garner:2014kna}.}. 
This is phrased as a statement about correlation functions.
To see this, consider loops constructed from a single complex adjoint matrix $Z$.
In terms of the normalized traces $O_J\equiv {\rm Tr}(Z^J)/\sqrt{JN^J}$ we have
\bea
\langle O_J^\dagger (x_1)O_J(x_2)\rangle &=&{1\over |x_1-x_2|^{2J}}+O\left({J^2\over N}\right)\cr\cr\cr
\langle O_{J_1+J_2}^\dagger (x_1)
O_{J_1}(x_2)O_{J_2}(x_2)\rangle &=& {\sqrt{J_1 J_2(J_1+J_2)}\over N|x_1-x_2|^{2J_1+2J_2}}+...\cr\cr
&\to& 0\qquad {\rm as}\qquad N\to\infty
\eea
The two point function of single traces is of order $1$, while the two point function of a double trace with a single
trace operator goes to zero.
We have considered mixing between single and double traces, but the conclusion is general: to mix different trace structures, we break color index loops to match traces structures and every time we break an index loop it costs a factor of $N$.
The fact that different trace structures do not mix in the planar limit is an important result, ultimately responsible for the existence of the spin chain language.
Indeed, the absence of mixing implies it is consistent to restrict to single trace operators and each single 
trace operator can be identified with a specific spin chain state.
We will derive a formula for the correlation functions of certain excitations of a (heavy) operator with an enormous 
$\sim N^2$ dimension in terms of the correlation functions of the planar limit.
As a consequence of this formula, we will see that simplifications of the planar limit encoded in correlation functions are then automatically present in correlation functions of certain excitations of the background.

We will make extensive use of the two point function of the restricted Schur polynomial, given by\cite{Bhattacharyya:2008rb}
\bea
\langle\chi_{R,(r,s)\alpha\beta}(Z,Y)\chi_{T,(t,u)\delta\gamma}(Z,Y)^\dagger\rangle
=\delta_{RS}\delta_{rt}\delta_{su}\delta_{\alpha\delta}\delta_{\beta\gamma} 
{f_R {\rm hooks}_R\over {\rm hooks}_r{\rm hooks}_s}
\eea
In the above formula $f_R$ stands for the product of factors of Young diagram $R$, while ${\rm hooks}_R$ stands
for the product of hook lengths of Young diagram $R$.
This result is exact for the free field theory, i.e. all ribbon diagrams have been summed.
Thus, the above formula is reliable for correlators of operators regardless of their dimension.
This is why its useful to express our computations in the restricted Schur polynomial language: we can tackle both the planar correlators (with dimension $\le O(\sqrt{N})$) and correlators in the background of a heavy operator (with dimension of $O(N^2)$) using a single formalism.

The computation of correlation functions most useful for our goals, starts by expressing the operators of
interest as linear combinations of restricted Schur polynomials.
This is always possible because the restricted Schur polynomials furnish a basis for the local gauge invariant operators of the theory.
An arbitrary operator $O_A$ 
\bea
O_A={\rm Tr}(\sigma Y^{\otimes m}\otimes Z^{\otimes n})
=Y^{i_1}_{i_{\sigma (1)}}\cdots Y^{i_m}_{i_{\sigma (m)}}
Z^{i_{m+1}}_{i_{\sigma (m+1)}}\cdots Z^{i_{n+m}}_{i_{\sigma (n+m)}}\label{genT}
\eea
can be written as a linear combination of restricted Schur polynomials as follows 
\bea
O_A = \sum_{R,r,s,\alpha,\beta} a^{(A)}_{R,(r,s),\alpha,\beta} \chi_{R,(r,s)\alpha\beta}(Z,Y,X,\cdots)\label{CS}
\eea
By changing the permutation $\sigma$ appearing in (\ref{genT}) we can obtain any desired multi trace structure. 
Taking linear combinations of these terms, we can easily construct, for example, the operators that would map into
the states of the spin chain.
Explicit formulas for the coefficients are known
\bea
{\rm Tr} (\sigma Z^{\otimes n}Y^{\otimes m})=\sum_{T,(t,u)\alpha\beta}
{d_T n! m!\over d_t d_u (n+m)!}\chi_{T,(t,u)\alpha\beta}(\sigma^{-1})
\chi_{T,(t,u)\beta\alpha}(Z,Y)\label{FTfRS}
\eea
We will not however need the precise values of the $a^{(A)}_{R,\{ r\}, \alpha}$.
Formula (\ref{FTfRS}) does however make it clear that these coefficients are symmetric group data and consequently, they are 
independent of $N$. 
Using the known two point function for the restricted Schur polynomial, we find in the free field theory, that
\bea
\langle O_A (x_1)O_B(x_2)^\dagger\rangle =
\sum_{R,r,s, \alpha} {a^{(A)}_{R,(r,s),\alpha,\beta} a^{(B)*}_{R,(r,s),\alpha,\beta}{\rm hooks}_R f_R\over
{\rm hooks}_r {\rm hooks}_s}{1\over |x_1-x_2|^{2J}}
\eea
The above result is exact and its an ingredient in the proof of the identity relating planar correlation functions of ${\cal N}=4$
super Yang-Mills theory to the correlations functions of the emergent gauge theories that arise in large $N$ but non-planar limits.  The planar approximation to the correlation function in free field theory is obtained by truncating the above exact result to its leading term in a large $N$ expansion.

Up to now we have focused on operators constructed using only the $Z$ and $Y$ fields.
The most general operator will be constructed from adjoint scalars, adjoint fermions or covariant derivatives of these fields.
The construction of restricted Schur polynomials with an arbitrary number of species of adjoint scalars and an arbitrary
number of species of adjoint fermions was given in \cite{Koch:2012sf}.
The construction of restricted Schur polynomials using covariant derivatives has been described in \cite{deMelloKoch:2011vn}.
Each power of the covariant derivative $D_\mu^p Z$ must be treated as a new species of field. 
If the operator we consider is constructed using a total of $k$ species of fields, then the restricted Schur polynomial
becomes $\chi_{R,\{r\},\alpha\beta}$, with $\{ r\}$ a collection of $k$ Young diagrams, one for each species.
If we use $n_i$ fields of species $i$ the corresponding Young diagram $r_i$ has $n_i$ boxes.
Young diagram $r_1$ corresponds to the $Z$ field.
Young diagram $R$ has $n_1+n_2+...+n_k$ boxes.
The additional labels contained in $\alpha$ and $\beta$ are again discrete labels distinguishing operators that carry 
the same $R,\{ r\}$ labels.
The formulas we have given above now generalize as follows
\bea
\langle\chi_{R,\{r\}\alpha\beta}(Z,Y)\chi_{T,\{t\}\delta\gamma}(Z,Y)^\dagger\rangle
=\delta_{RS}\delta_{\{r\},\{t\}}\delta_{\alpha\delta}\delta_{\beta\gamma} 
{f_R {\rm hooks}_R\over \prod_r {\rm hooks}_r}
\eea
\bea
O_A = \sum_{R,r,s,\alpha,\beta} a^{(A)}_{R,\{r_1,r_2,\cdots\},\alpha,\beta} 
\chi_{R,\{r_1,r_2,\cdots\}\alpha\beta}(Z,Y,X,\cdots)\label{CSII}
\eea
and
\bea
\langle O_A (x_1)O_B(x_2)^\dagger\rangle =
\sum_{R,r,s, \alpha} {a^{(A)}_{R,\{r\},\alpha,\beta} a^{(B)*}_{R,\{r\},\alpha,\beta}{\rm hooks}_R f_R\over
\prod_r {\rm hooks}_r}{1\over |x_1-x_2|^{2J}}
\eea
In the above formulas, $\delta_{\{r\},\{t\}}$ is 1 if the complete ordered sets of Young diagrams $\{r\}$ and $\{t\}$ are
equal, and it is zero otherwise. The planar approximation is again obtained by truncating to the leading term in a large $N$ expansion. This completes our discussion of the planar correlation functions.

\subsection{Exitations of an LLM Geometry}\label{NPCorr}

\begin{center}
\begin{figure}[!h]
\centering
\includegraphics[width=0.7\textwidth]{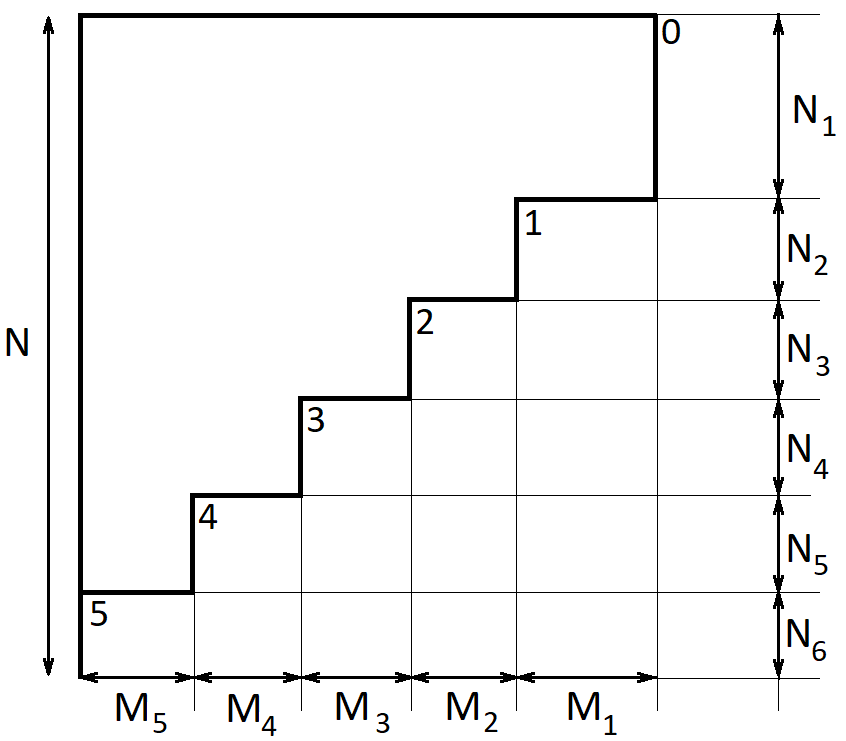}
\caption{A possible label $B$ for a Schur polynomial describing an LLM background. Note that
$\sum_{i=1}^6 N_i=N$.}
\label{LLM}
\end{figure}
\end{center}

The LLM geometries that we consider are described by Schur polynomials $\chi_B(Z)$ of the complex matrix $Z$
labeled by a Young diagram $B$ with $O(N^2)$ boxes and $O(1)$ outward pointing corners.
An example of a possible Young diagram $B$, with 5 outward pointing corners is shown in Figure \ref{LLM}.

\begin{center}
\begin{figure}[!h]
\centering
\includegraphics[width=0.52\textwidth]{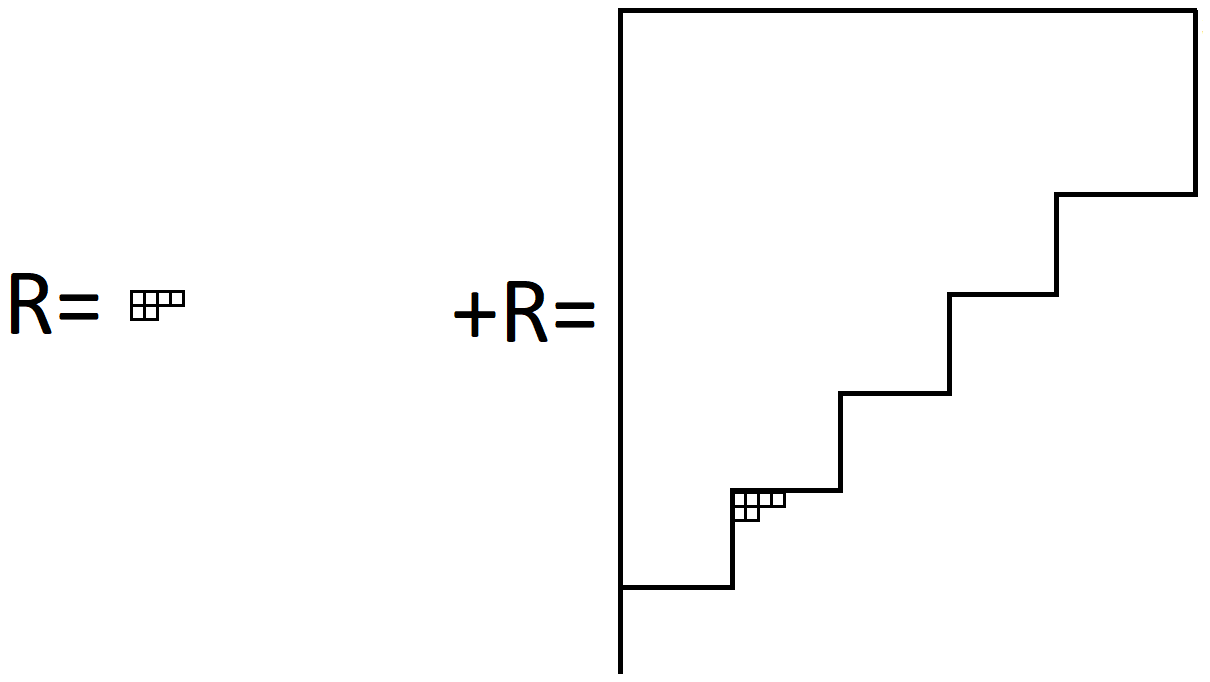}
\caption{To obtain $+R$ from $R$ we stack $R$ at one of the inward pointing corners of $B$.}
\label{ChangeR}
\end{figure}
\end{center}

All of the horizontal edges $M_i$, and vertical edges $N_i$ have a length of $O(N)$.
Excitations are obtained by adding $J=O(\sqrt{N})$ boxes to $B$.
These new boxes could be stacked at any of the inward pointing corners, below or to the right\footnote{We could 
also create excitations by eroding the outward pointing corners.
We will not study these excitations here.} of $B$.
The possible locations for the new boxes are labeled 0 to 5 in Figure \ref{LLM}.
We will distinguish between excitations constructed by adding all extra boxes at a single inward pointing corner ({\it localized} excitations) and excitations constructed by adding extra boxes at more than one corner ({\it delocalized} excitations).
In the free field theory, thanks to the fact that the two point function of the restricted Schur polynomial is diagonal
in all of its labels, the local and delocalized excitations are orthogonal\footnote{When we make this comment we have the
operator/state correspondence of the CFT in mind. According to the correspondence, the inner product of two states is
related to the correlators of the corresponding operators.}.  
Denote the Hilbert space of small fluctuations about the LLM geometry by ${\cal H}_{\rm CFT;LLM}$.
This Hilbert space can be decomposed as a direct sum as follows
\bea
   {\cal H}_{\rm CFT;LLM}={\cal H}_{\rm CFT;Local}\bigoplus {\cal H}_{\rm CFT; Delocalized}
\eea
Our study will focus on the local excitations.
The Hilbert space of local excitations can further be refined as a direct sum of subspaces, one for each corner of the
background Young diagram
\bea
   {\cal H}_{\rm CFT;Local}=\bigoplus_i {\cal H}^{(i)}_{\rm CFT}
\eea
where $i$ runs over inward pointing corners with the understanding that below or to the right\footnote{The locations labeled 0 and 5 in Figure \ref{LLM}.} of $B$ count as corners.
Each factor ${\cal H}^{(i)}_{\rm CFT}$ in the above sum is the Hilbert space of an emergent gauge theory and is isomorphic to the space of local operators in the planar limit of the original CFT, as we now explain.
We do this by giving the bijection between operators of dimension $J$ with $J^2\ll N$ and operators in 
${\cal H}^{(i)}_{\rm CFT}$.
The bijection maps the operator given in (\ref{CSII}) above into
\bea
O^{(B)}_A = \sum_{R,r,s,\alpha,\beta} a^{(A)}_{R,\{r_1,r_2,\cdots\},\alpha,\beta} 
\chi_{+R,\{+r_1,r_2,\cdots\}\alpha\beta}(Z,Y,X,\cdots)\label{CSIII}
\eea
The coefficients of the expansion appearing in (\ref{CSII}) are identical to the coefficients appearing in (\ref{CSIII}).
It is only the $R$ and $r_1$ labels in the restricted Schur polynomials in (\ref{CSII}) and (\ref{CSIII}) that have changed.
The Young diagram $+R$ is obtained by stacking $R$ at the $i$th corner of $B$ and similarly, the
Young diagram $+r$ is obtained by stacking $r$ at the $i$th corner of $B$.
For an example of how this works, see Figure \ref{ChangeR}.

This mapping is a bijection. Operators with distinct labels are orthogonal. Operators with distinct labels before the map have distinct labels after the map. Thus, the mapping is injective.
Any operator with a bare dimension $J$ and $J^2\ll N$ can be mapped to an excitation of the background $B$.
What is important here is that, since each edge of the Young diagram has a length of order $N$, there is no danger
that when we stack $R$ it will not fit onto the corner.
Of course, the converse is also true: any excitation of the background can be mapped to an operator of dimension
$J$ by deleting the boxes in $+R$ and $+r_1$ which belong to $B$.
Thus, the map is surjective.
This demonstrates that our mapping is a bijection. 

In the remainder of this section we will argue that the correlation functions of operators that are in bijection are related 
in a particularly simple way, in the large $N$ limit.
We would like to normalize our correlators so that
\bea
\langle 1\rangle_B =1
\eea
We know that $1$ maps into $\chi_B(Z)$ and that
\bea
\langle\chi_B(Z)\chi_B(Z)^\dagger\rangle =f_B{1\over |x_1-x_2|^{2|B|}}
\eea
where $|B|$ is the free field dimension of $\chi_B(Z)$. Consequently we will include an extra factor of $|x_1-x_2|^{2|B|}f_B^{-1}$ to ensure that our correlators are correctly normalized
\bea
    \langle\cdots\rangle_B ={\langle\cdots\rangle\over f_B}|x_1-x_2|^{2|B|}
\eea
Using the two point function of the restricted Schur polynomial, we obtain the following result
\bea
\langle O^{(B)}_A (x_1)O^{(B)}_B(x_2)^\dagger\rangle_B =
\sum_{R,r,s, \alpha} {a^{(A)}_{R,(r,s),\alpha,\beta}\, a^{(B)*}_{R,(r,s),\alpha,\beta}
\,{\rm hooks}_{+R}\, f_{+R}\over f_B\, {\rm hooks}_{+r_1} \prod_{i>2}{\rm hooks}_{r_i}}
{1\over |x_1-x_2|^{2J}}
\eea
Assume that we have a total of $C$ outward pointing corners and further that our localized excitation
is stacked in the $i$th corner.
Applying the identity (\ref{HIdentity}) we find

\bea
\langle O^{(B)}_A (x_1)O^{(B)}_B(x_2)^\dagger\rangle_B =
(\eta_B)^{n_I}\sum_{R,r,s, \alpha} {a^{(A)}_{R,(r,s),\alpha,\beta}\, a^{(B)*}_{R,(r,s),\alpha,\beta}\,
{\rm hooks}_{R}\over\prod_{i}{\rm hooks}_{r_i}}{f_{+R}\over f_B}{1\over |x_1-x_2|^{2J}}
\left( 1+O\left({1\over N}\right)\right)\cr
\eea
(\ref{HIdentity}) is not exact - it includes ${1\over N}$ corrections and this is the only source of ${1\over N}$ corrections in our final result.
We have assumed\footnote{This assumption is not necessary. By rescaling each impurity field by $\sqrt{\eta_B}$ we could remove the $\eta_B$ dependence in these formulas.} that every term in the sum has the same total number 
of fields and the same number of $Z$ fields, i.e. that each term has the same value for $|R|-|r_1|\equiv n_I$.
The subscript $I$ on $n_I$ stands for ``impurity'' since its common to refer to fields in our excitation that are not $Z$
fields as impurities.
We would now like to compare this to the result that we obtained for the planar correlators, which is
\bea
\langle O_A (x_1)O_B(x_2)^\dagger\rangle &=&
\sum_{R,r,s, \alpha} {a^{(A)}_{R,(r,s),\alpha,\beta} a^{(B)*}_{R,(r,s),\alpha,\beta}{\rm hooks}_R f_R\over
{\rm hooks}_r {\rm hooks}_s}{1\over |x_1-x_2|^{2J}}\cr
&\equiv&F_{AB}(N){1\over |x_1-x_2|^{2J}}
\eea
The two results are nearly identical.
The only difference, apart from the overall factor $(\eta_B)^{n_I}$, is that $f_R$ in the planar result is replaced by ${f_{+R}\over f_B}$ in the emergent gauge theory result.
Now, recall that $f_R$ is the product of factors in Young diagram $R$ and that a box in row $j$ and column $k$
has factor $N-j+k$.
Consequently
\bea
   f_R(N)=\prod_{(j,k)\in R} (N-j+k)
\eea
In the ratio ${f_{+R}\over f_B}$ factors of boxes that are common to $+R$ and $B$ cancel.
After performing these cancellations we find
\bea
  {f_{+R}\over f_B}=f_R(N_{\rm eff})=\prod_{(j,k)\in R} (N_{\rm eff}-j+k)
\eea
where
\bea
   N_{\rm eff}=N-\sum_{a=1}^i N_a+\sum_{b=i+1}^C M_b
\eea
This last formula is explained in Figure \ref{LLM} and Appendix \ref{Ratf}.
$N_{\rm eff}$ is the factor of the first excitation box added to the background Young diagram.
Finally, recalling that the only source of $N$ dependence is in $f_R$ (for the planar correlators) or ${f_{+R}\over f_B}$
(for the emergent gauge theory correlators) we finally obtain
\bea
\langle O_A (x_1)O_B(x_2)^\dagger\rangle &=&F_{AB}(N){1\over |x_1-x_2|^{2J}}\cr
\langle O_A (x_1)O_B(x_2)^\dagger\rangle_B &=& F_{AB}(N_{\rm eff}){(\eta_B)^{n_I}\over |x_1-x_2|^{2J}}
\left( 1+O\left({1\over N}\right)\right)\label{CoolR}
\eea
This demonstrates a remarkable relationship between correlators in the planar and non-planar limits.

This result has a number of immediate applications.
As we have stressed above, the fact that operators with different trace structures don't mix in the planar limit is a statement 
about correlators.
This no-mixing result allows focus on the single trace operators which is needed to develop the spin chain connection for the planar CFT.
Our result (\ref{CoolR}) immediately implies that operators that are the image of operators with different trace structures,
will not mix. 
Thus, we too can focus on the image of single trace operators and then develop a spin chain description of the planar
limit of the emergent gauge theory.
States of the spin chain that were identified with a given operator in the planar limit will now be identified with the
image of the same operator.

Three point functions of single trace operators are suppressed in the planar limit of the original CFT. Is there a similar statement for three point functions of single trace operators in the emergent gauge theory? In any Poincare invariant CFT the spacetime dependence of the three point function is fixed by conformal invariance. We can thus simply factor this dependence out and consider the problem of the combinatorics of the Wick contractions. This is also a complicated problem, but for some well chosen examples it can be solved. Consider the following correlator
\bea
\langle \Tr(Z^{n_1}Y^{n_2}\cdots)\Tr(Z^{m_1}Y^{m_2}\cdots)\Tr(Z^{\dagger p_1}Y^{\dagger p_2}\cdots)\rangle
\eea  
The Wick contractions are all between the first trace and the third trace, and between the second trace and the third trace.
In particular, there are no contractions between the first and second traces. For the combinatorics of the Wick contractions, we can treat the double trace $\Tr(Z^{n_1}Y^{n_2}\cdots)\Tr(Z^{m_1}Y^{m_2}\cdots)$ as a single operator and apply the bijection and treat $\Tr(Z^{\dagger p_1}Y^{\dagger p_2}\cdots)$ as a single operator and apply the bijection. Thus, we have
reduced the computation to a two point function. This two point correlator correctly sums the contractions between the three traces with each other and with the background.
The result (\ref{CoolR}) then implies that this correlator, which is giving the three point function, is suppressed in the planar limit of the emergent gauge theory.
Since OPE coefficients are read from three point functions, the OPE coefficients vanish in the planar limit of both
${\cal N}=4$ super Yang-Mills and the emergent gauge theory. We have proved this in the free field theory, for a specific class of correlators. We {\it conjecture} that it holds quite generally and continues to hold when interactions are turned on.
The usual suppression holds because we need to break index loops (which costs $N^{-1}$ for each loop we break) to find a non-zero correlator between three single traces. This does not rely on any detailed structure of the interaction and is quite generally true for a matrix model. Of course, this is one point in our analysis that could be improved.

Our argument in this section considers only the local operators. 
One might wonder if mixing between different trace structures of delocalized operators is also suppressed or not.
In this case the argument is more involved.
It is unlikely that there is a simple relationship between correlators of delocalized operator and correlators
computed in the planar limit.
Explicit computations using concrete examples support the conclusion that again, different trace structures don't mix.
See Appendix \ref{deloc} for a discussion of this point.

To summarize, we have arrived at a rather detailed picture of the structure of the Hilbert space.
We have decomposed the Hilbert space of excitations of the LLM geometry into a direct sum
\bea
   {\cal H}_{\rm CFT;LLM}=\left(\bigoplus_i {\cal H}^{(i)}_{\rm CFT}\right)\bigoplus {\cal H}_{\rm CFT; Delocalized}
\eea
Restricted Schur polynomials are orthogonal if their labels don't match.
This immediately implies that in the free field theory operators belonging to different Hilbert spaces in the above sum have
vanishing two point functions and hence that the corresponding subspaces are orthogonal.
We have further argued that each subspace can be decomposed into a direct sum of orthogonal components, with
each component collecting operators of a definite ``trace structure''. Here the trace structure is read from the preimage of the operator under the bijection (\ref{CSIII}).
At large $N$ these different trace structures do not mix.

Our study has focused on the free field theory. Of course, the bijection we have defined holds for any coupling. The free field limit has been used to obtain the relationship between correlators of operators and correlators of their images. It is this discussion that we will extend to weak coupling in the next section.

\section{Weak Coupling CFT}

We expect that the gravitational physics dual to the CFT is coded into the large $N$ correlators.
Consequently, it is attractive if we can find relationships between correlators of the planar limit and correlators in 
the background of a heavy operator.
In the previous section we have exhibited relationships of this type, all in the free limit of the CFT.
We expect the dual gravitational description is simplest when the CFT is strongly coupled.
It is natural to ask if the simple relations between correlation functions exhibited in the free theory
survive when interactions are added.
Answering this question is the goal of the current section.
We start with a careful discussion of the one loop dilatation operator, which develops the relation between correlators
at one loop.
This argument also gives insight into why the relationship we have uncovered between correlators holds even
when higher loop corrections are included.

The article \cite{Koch:2016jnm} argued that matrix elements of the planar dilation operator are identical to matrix elements of the dilatation operator computed using local excitations, localized at corner\footnote{These operators belong to ${\cal H}_{\rm CFT}^{(i)}$} $i$ of the Young diagram for the LLM geometry, after replacing $\lambda=g_{YM}^2 N$ by $\lambda_{\rm eff}=g_{YM}^2 N_{\rm eff}$  where $N_{\rm eff}$ is the factor of the first box added to corner $i$.
This again amounts to replacing $N\to N_{\rm eff}$ so it is the rule we derived in Section \ref{NPCorr}!
We will revisit this argument below adding two new improvements
\begin{itemize}
\item[1.] By carefully tracking what is background independent and what is not we will develop a much simpler technical analysis.
\item[2.] We will phrase the result using the bijection we developed in Section \ref{NPCorr}. The advantage of the rephrasing is that it supports the conclusion that the planar limit of the emergent gauge theory is planar ${\cal N}=4$ super Yang-Mills.  
\end{itemize}
The final result is remarkable: in the large $N$ but non-planar limit we need to sum a huge set of Feynman diagrams.
The net effect of summing the huge set of non-planar diagrams, is a simple rescaling of the 't Hooft coupling.
This is in complete harmony with the physical argument we developed in the introduction.

The fact that we simply need to rescale the 't Hooft coupling has far reaching consequences: since the dilatation operator in the planar CFT matches the Hamiltonian of an integrable spin chain, we know that the dilatation operator describing the anomalous dimensions of the emergent gauge theory will also match the integrable spin chain.
As long as the dilatation operator does not mix operators that belong to the Hilbert space ${\cal H}_{\rm CFT}^{(i)}$ with operators that don't belong to this space, we conclude that there are integrable subsectors in the large $N$ but non-planar limit we consider. Demonstrating the absence of this mixing is one of the main goals of this section.

Before turning to a detailed technical analysis we will briefly review the evidence supporting the above result.
It implies that the anomalous dimensions of the operators of the planar emergent gauge theory are determined in terms of the corresponding dimensions computed in the planar CFT.
Explicit computations of anomalous dimensions of the emergent CFT, when developed in a perturbative expansion, confirm this prediction both in the weak coupling CFT and at strong coupling using the dual string theory\cite{Koch:2015pga,Koch:2016jnm}.
Using the $su(2|2)$ symmetry enjoyed by the $su(3|2)$ subspace of local excitations, the two magnon $S$-matrix has been
determined and it agrees up to two loops with a weak coupling computation performed in the CFT\cite{deMelloKoch:2018tlb}.
The first finite size corrections to both the magnon and the dyonic magnon have been computed by constructing solutions to the Nambu-Goto action that carry finite angular momentum.
These computations\cite{deMelloKoch:2018tlb} again show that the net affect of the background is a scaling of 
the 't Hooft coupling.
This constitutes strong coupling evidence for our result.
Since these corrections are sensitive to the overall phase of the $S$-matrix, which is not determined by kinematics
(i.e. the $su(2|2)^2$ symmetry of the theory), this is a non-trivial test.
Finally, strings spinning on the three sphere that belongs to AdS$_5$ have been considered in \cite{Kim:2018gwx}.
These strings are dual to operators belonging to the $SL(2)$ sector of the gauge theory.
Once again, the net affect of the background is a scaling of the 't Hooft coupling as predicted\cite{Kim:2018gwx}.

In the subsection that follows we revisit the analysis of \cite{Koch:2016jnm}, phrasing things in terms of the
bijection of section \ref{NPCorr} and paying attention the background dependence of the various ingredients in the analysis.
This significantly simplifies the original analysis.
We pay careful attention to operator mixing, to give evidence supporting the conclusion that the integrable subsectors are decoupled at large $N$.
This closes an important hole in the analysis of \cite{Koch:2016jnm}.
Finally, we consider how the one loop discussion generalizes when we include higher loops.

\subsection{One Loop Mixing of Local Operators}

From now on we normalize the two point function of our operators to 1.
To simplify the discussion again focus on operators constructed using only $Z$ and $Y$ fields.
It is a simple generalization to include more fields.
Consider the mixing between two restricted Schur polynomials, $O_{+R,(+r,s)\mu_1\mu_2}(Z,Y)$ and
$O_{+T,(+t,u)\nu_1\nu_2}(Z,Y)$.
The capital letter $O$ for the restricted Schur polynomial instead of the $\chi$ stresses the fact we are considering normalized operators
\bea
\langle O_{+R,(+r,s)\mu_1\mu_2}(Z,Y)^\dagger O_{+T,(+t,u)\nu_1\nu_2}(Z,Y)\rangle
=\delta_{+R,+T}\delta_{+r,+t}\delta_{su}\delta_{\mu_1\nu_1}\delta_{\mu_2\nu_2}
\eea
These operators are the image under the bijection of $O_{R,(r,s)\mu_1\mu_2}(Z,Y)$ and $O_{T,(t,u)\nu_1\nu_2}(Z,Y)$.
The operators $O_{+R,(+r,s)\mu_1\mu_2}(Z,Y)$ provide a basis for ${\cal H}_{\rm CFT}^{(i)}$.
The starting point of our analysis is the one loop dilatation operator in this basis\cite{DeComarmond:2010ie}
\bea
  DO_{+R,(+r,s)\mu_1\mu_2}(Z,Y)
=\sum_{T,(t,u)\nu_1\nu_2}N_{+R,(+r,s)\mu_1\mu_2;+T,(+t,u)\nu_1\nu_2}O_{+T,(+t,u)\nu_1\nu_2}(Z,Y)
\label{dilact}
\eea
where
\bea
N_{+R,(+r,s)\mu_1\mu_2;+T,(+t,u)\nu_1\nu_2}
=-{g_{YM}^2\over 8\pi^2}\sum_{+R'}
{c_{+R,+R'} d_{+T} n m \over d_{+R'}d_{+t} d_u (n+m)}
\sqrt{f_{+T} {\rm hooks}_{+T}{\rm hooks}_{+r}{\rm hooks}_s\over f_{+R}{\rm hooks}_{+R}
{\rm hooks}_{+t}{\rm hooks}_u}\cr
{\rm Tr}\left(\big[ (1,m+1),P_{+R,(+r,s)\mu_1\mu_2}\big]I_{+R'+T'}
\big[(1,m+1),P_{+T,(+t,u)\nu_2\nu_1}\big]I_{+T' +R'}\right)\cr
\label{matelem}
\eea
In the above expression Young diagram $+R'$ is obtained by dropping one box from $+R$ and $c_{+R,+R'}$
is the factor of the box that is dropped.
Also, $d_r$ is the dimension of symmetric group irreducible representation $r$.
Use $n$ to denote the total number of $Z$ fields in $O_{+R,(+r,s)\mu_1\mu_2}$ and $n_B$ to denote the 
number of $Z$ fields in the background. Also, $n_Z$ denotes the number of $Z$ fields in $O_{R,(r,s)\mu_1\mu_2}$
and $m$ denotes the number of $Y$ fields. We have $n=n_B+n_Z$.
The above result (\ref{dilact}),(\ref{matelem}) was derived using the convention that the $Y$ fields occupy slots 1 to $m$ exactly as shown in
(\ref{restrictedschur}).
In the standard tableau labeling of the states in $+R$, the $Y$'s would be associated to the boxes labeled 1 to $m$.
This result is the exact one loop result - we have not made use of any of the simplifications that come from taking $N\to\infty$.
Notice that the $N$ dependence of the matrix elements appears in $c_{+R,+R'}$, $f_{+R}$ or $f_{+T}$.
This immediately implies that we will again have a dependence on $N_{\rm eff}$ and not on $N$.

To proceed further, begin by discussing the intertwining map $P_{+R,(+r,s)\mu_1\mu_2}$.
Our goal is to give a careful argument concluding that $P_{+R,(+r,s)\mu_1\mu_2}$ is background independent.
This map acts within a direct sum of the carrier space of $+T$ and the carrier space of $+R$.
It gives zero on $+T$ and projects the row and column labels of the $+R$ subspace to an $(r,s)$ irreducible representation of $S_n\times S_m$.
Our convention is that the first boxes removed are associated to $Y$. This projection operator
simply has to assemble these boxes into an irreducible representation $s$ of $S_m$.
The remaining boxes are already in $+r$.
Thus, the projection operator is
\bea
 P_{+R,(+r,s)}={1\over m!}\sum_{\sigma\in S_m}\chi_s(\sigma)\Gamma_{+R}(\sigma)
\eea
In writing the above projection operator it is understood that we are acting in the subspace of $+R$ in which states are labeled by standard tableau such that labels 1,...,$m$ only fill boxes that belong to $+R$ and not to $+r$.
This is the subspace in which the remaining boxes are already in $+r$.
To get the intertwining map, restrict the above row and column labels.
The key point is that the projection operator acts only on boxes associated to the $Y$ fields.
Restricting indices to get the intertwining map will not change this so that the intertwining map
$P_{+R,(+r,s)\alpha\beta}$ only has a nontrivial action on the $Y$ boxes, that is, on the boxes that are removed
from $+R$ to get $+r$.
With the discussion of Section \ref{bckindep} in mind, its clear that $P_{+R,(+r,s)\alpha\beta}$ is background independent.

To evaluate the matrix elements of the dilatation operator, we need to perform the following trace
\bea
{\rm Tr}\left(\big[ (1,m+1),P_{+R,(+r,s)\mu_1\mu_2}\big]I_{+R'+T'}
\big[(1,m+1),P_{+T,(+t,u)\nu_2\nu_1}\big]I_{+T' +R'}\right)\label{Trtodo}
\eea
The intertwining maps $I_{+R'+T'}$ and $I_{+T' +R'}$ map from the subspace $+R'$ obtained by dropping a single
box from $+R$, to the subspace $+T'$ obtained by dropping a single box from $+T$.
As a result, these maps act only on the box in the standard tableau labeled 1 which is associated to a $Y$ and hence these maps are background independent.
The results of Section \ref{bckindep} imply that the above trace is background independent.
Lets pursue this further in our current example.
The intertwining maps $P_{+R,(+r,s)\mu_1\mu_2}$ and $P_{+T,(+t,u)\nu_2\nu_1}$ act only on the boxes labeled
1 to $m$ - all $Y$ boxes, and the permutation $(1,m+1)$ acts only on boxes labeled $1$ or $m+1$. 
One is a $Y$ box, one is a $Z$ box and both belong to the excitation.
Consequently, in the above trace the very vast majority of boxes - those with labels $> m+1$ and there are $O(N^2)$ of them - are simply spectators and can be traced over.
Recall that we are focusing on operators that belong to a given emergent gauge theory.
The non-trivial structure of the matrix elements is determined by the Young diagrams $R$, $r$ and $s$ and it will agree with the non-trivial structure of the planar matrix elements - this is the background independence.
The only difference between the planar result for the trace and what we consider above, is that the sum over the inert boxes produces a factor $d_{+r_i'}$ where $+r_i'$ is obtained by dropping a box from row $i$ of $r$ in $+r$ while in the planar case we get a factor of $d_{r_i'}$.
If we now consider mixing with operators outside of the emergent gauge theory, in principle we could drop a box from $+r$ at any location - even a corner that is distinct from where our excitation is located. These matrix elements arise when there is mixing with states that don't belong to the integrable subsector. We will consider these corrections in detail in the next section. Our conclusion is that these matrix elements vanish at large $N$. Using this result, we can restrict to mixing between operators that belong to the planar limit of the emergent gauge theory. Consequently, the bijection of section \ref{NPCorr} relates these operators to two operators, $O_{R,(r,s)\mu_1\mu_2}(Z,Y)$ and $O_{T,(t,u)\nu_1\nu_2}(Z,Y)$ defined in the planar CFT.
We will now derive a relationship between the matrix elements for mixing $O_{+R,(+r,s)\mu_1\mu_2}(Z,Y)$
and $O_{+T,(+t,u)\nu_1\nu_2}(Z,Y)$ and those for mixing $O_{R,(r,s)\mu_1\mu_2}(Z,Y)$ and $O_{T,(t,u)\nu_1\nu_2}(Z,Y)$. This extends the free field theory relationship between correlators obtained in Section \ref{NPCorr}, to one loop.
The argument is\footnote{Recall that $d_r=n!/{\rm hooks}_r$ for any irrep $r$ of $S_n$.
In what follows ${\rm Tr}_i$ indicates that we have traced over $r_i$ and ${\rm Tr}_{+i}$ indicates that we have traced over $+r_i$.}
\bea
&&-{g_{YM}^2\over 8\pi^2}\sum_{+R'}
{c_{+R,+R'} d_{+T} n m \over d_{+R'}d_{+t} d_u (n+m)}
\sqrt{f_{+T} {\rm hooks}_{+T}{\rm hooks}_{+r}{\rm hooks}_s\over f_{+R}{\rm hooks}_{+R}
{\rm hooks}_{+t}{\rm hooks}_u}\cr
&&\qquad {\rm Tr}\left(\big[ (1,m+1),P_{+R,(+r,s)\mu_1\mu_2}\big]I_{+R'+T'}
\big[(1,m+1),P_{+T,(+t,u)\nu_2\nu_1}\big]I_{+T' +R'}\right)\cr\cr
&&=
-{g_{YM}^2\over 8\pi^2}\sum_{+R'}\sum_i
{c_{+R,+R'} m \over d_u}
{\sqrt{{\rm hooks}_{+r}{\rm hooks}_{+t}}\over{\rm hooks_{+r_i'}}}
{{\rm hooks}_{+R'}\over\sqrt{{\rm hooks}_{+T}{\rm hooks}_{+R}}}
\sqrt{f_{+T} {\rm hooks}_s\over f_{+R}{\rm hooks}_u}
\cr
&&\qquad {\rm Tr}_{+i}\left(\big[ (1,m+1),P_{+R,(+r,s)\mu_1\mu_2}\big]I_{+R'+T'}
\big[(1,m+1),P_{+T,(+t,u)\nu_2\nu_1}\big]I_{+T' +R'}\right)\cr\cr
&&=-{g_{YM}^2\over 8\pi^2}\sum_{+R'}\sum_i
{c_{+R,+R'} m \over d_u}
{\sqrt{{\rm hooks}_{r}{\rm hooks}_{t}}\over{\rm hooks_{r_i'}}}
{{\rm hooks}_{R'}\over\sqrt{{\rm hooks}_{T}{\rm hooks}_{R}}}
\sqrt{f_{+T} {\rm hooks}_s\over f_{+R}{\rm hooks}_u}
\cr
&&\qquad {\rm Tr}_i\left(\big[ (1,m+1),P_{+R,(+r,s)\mu_1\mu_2}\big]I_{+R'+T'}
\big[(1,m+1),P_{+T,(+t,u)\nu_2\nu_1}\big]I_{+T' +R'}\right)\cr\cr
&&=-{g_{YM}^2\over 8\pi^2}\sum_{+R'}\sum_i
{c_{+R,+R'} m \over d_u}
{\sqrt{{\rm hooks}_{r}{\rm hooks}_{t}}\over{\rm hooks_{r_i'}}}
{{\rm hooks}_{R'}\over\sqrt{{\rm hooks}_{T}{\rm hooks}_{R}}}
\sqrt{f_{+T} {\rm hooks}_s\over f_{+R}{\rm hooks}_u}
\cr
&&\qquad {\rm Tr}_i\left(\big[ (1,m+1),P_{R,(r,s)\mu_1\mu_2}\big]I_{R'T'}
\big[(1,m+1),P_{T,(t,u)\nu_2\nu_1}\big]I_{T'R'}\right)\cr\cr
&&=-{g_{YM}^2\over 8\pi^2}\sum_{R'}
{c_{+R,+R'} d_{T} n_Z m \over d_{R'}d_{t} d_u (n_Z+m)}
\sqrt{f_{+T} {\rm hooks}_{T}{\rm hooks}_{r}{\rm hooks}_s\over f_{+R}{\rm hooks}_{R}
{\rm hooks}_{t}{\rm hooks}_u}\cr
&&\qquad {\rm Tr}\left(\big[ (1,m+1),P_{R,(r,s)\mu_1\mu_2}\big]I_{R'T'}
\big[(1,m+1),P_{T,(t,u)\nu_2\nu_1}\big]I_{T'R'}\right)
\eea
In moving to the third line above we have used the formula (\ref{HIdentity}) proved in Appendix \ref{ProveIdentity}.
This is the only step in the above computation that is not exact, but relies on the large $N$ limit.
Notice that the only difference between the last line above and the matrix elements of the dilatation operator
in the planar limit is that $N$ is replaced with $N_{\rm eff}$.
This is then a simple proof that at large $N$, the matrix elements of the one loop dilatation operator with respect to states of the emergent gauge theory are given by replacing $N\to N_{\rm eff}$ in the matrix elements of the planar dilatation operator, taken with respect to the preimages of these states.

How does this generalize to higher loops? The two loop dilatation operator has been considered in \cite{deMelloKoch:2012sv} and from that analysis it is clear what the general results are. The structure of the matrix elements are very similar to the form shown in (\ref{matelem}). One again lands up computing a trace. The same intertwining maps $P_{+R,(+r,s)\mu_1\mu_2}$ and $P_{+T,(+t,u)\nu_1\nu_2}$ appear in the trace. The maps $I_{R'T'},I_{T'R'}$ are replaced at $L$ loops by maps which map from a representation $R^{(L)}$ obtained by dropping $L$ boxes from $R$ to a representation $T^{(L)}$ obtained by dropping $L$ boxes from $T$. There are also again permutations that act on the boxes associated to the excitation. Finally, the trace is multiplied by the square root of the factors of the boxes dropped from $R$ and $T$. Arguing as we did above, its clear that the trace is background independent and the product of factors implies that the simple rule $N\to N_{\rm eff}$ again applies. These observations imply that our one loop conclusion goes through when higher loop corrections are included.

To summarize, we have found integrable subsectors in the large $N$ but non-planar limit that we are considering.
Each integrable subsector is an emergent gauge theory, with its own gauge group $U(N_{\rm eff})$.
To complete this discussion, in the next section we will consider the mixing between the integrable and non-integrable subsectors.

\subsection{Mixing with Delocalized Operators}\label{NBMix}

The operators that belong to the planar limit of a given emergent gauge theory are localized at a given corner and define an integrable subsector of the theory.
There are operators that are not localized at one corner - they straddle two or more corners.
If these delocalized operators mix with the localized operators they will almost certainly ruin integrability of the emergent gauge theory.
In this section we consider the mixing between localized and delocalized operators.
Our main result is that
\bea
\langle \phi | D | \psi\rangle =0 \qquad |\phi\rangle \in {\cal H}_{\rm CFT;Local}\quad |\psi\rangle\in 
{\cal H}_{\rm CFT;Delocalized}\label{bigMix}
\eea
at large $N$.

We make extensive of two basic observations.
First, in computing the matrix element (\ref{matelem}), it is clear that the reason why two different states
can have a non-zero matrix element, is because the permutation group element $(1,m+1)$ acts to change the
identity of the state.
It is thus important to have a good understanding of the action of this permutation on a standard tableau.
Since we are computing a trace which has the same value in any equivalent representation, we can carry this computation 
out in any convenient representation.
In what follows, we will use Young's orthogonal representation.
This representation is specified by giving the action of adjacent swaps  which are two cycles of the form $(i,i+1)$.
Given the matrices representing the complete set of adjacent swaps, it is easy to generate the rest of the group.
Let $|\psi\rangle$ denote a valid standard tableau and let $|\psi\rangle_{i\leftrightarrow i+1}$ denote the
state obtained from $|\psi\rangle$ by swapping $i$ and $i+1$. 
The content of the box labeled $i$, denoted $c(i)$ is given by $b-a$ if the box is in row $a$ and column $b$.
Our convention for the standard tableau labeling is spelled out in the following example
\bea
\young(54321)
\eea
The rule specifying the matrix representing the adjacent swap is
\bea
   (i,i+1)|\psi\rangle ={1\over c(i)-c(i+1)}|\psi\rangle 
+\sqrt{1-{1\over (c(i)-c(i+1))^2}}|\psi\rangle_{i\leftrightarrow i+1}\label{defrep}
\eea
If boxes $i$ and $i+1$ are located at different corners, the first term above is of order $N^{-1}$ and can be neglected
in the large $N$ limit while the coefficient of the second term is 1, to the same accuracy.

The second observation is a relationship between the loop order and the number of boxes that can differ in the Young
diagram labels of the operators that are mixing.
To add loop effects, we consider Feynman diagrams with a certain number of vertices included in the diagram. 
Contracting two fields in a restricted Schur polynomial with a vertex has the effect of setting the indices of two different
fields equal.
This Kronecker delta function restricts the sum over permutations in (\ref{restrictedschur}) from  $S_{n+m}$ to
$S_{n+m-1}$.
Two operators which begin as distinct representations of $S_{n+m}$ may well produce the same representation
of $S_{n+m-1}$.
For this to happen, their Young diagram labels must differ in the placement of at most one box.
This is manifest in the matrix element (\ref{matelem}), because the maps $I_{T'R'}$ which appear are
only non-zero if $T'$ (obtained by dropping one box from $T$) has the same shape as $R'$ (obtained by dropping one 
box from $R$).
At $L$ loops we have added $L$ vertices which lands up restricting the sum in (\ref{restrictedschur}) from  $S_{n+m}$ to
$S_{n+m-L}$.
In this case operators that differ by at most $L$ boxes will mix. 

As a warm up example, consider the mixing of localized operators that belong to different corners
\bea
\langle \phi | D | \psi\rangle =0 \qquad |\phi\rangle \in {\cal H}^{(i)}_{\rm CFT}\quad |\psi\rangle\in 
{\cal H}^{(j)}_{\rm CFT}\label{nomix}
\eea
with $j\ne i$.
This represents a mixing between states of two different planar emergent gauge theories, i.e. two distinct integrable subsectors.
For concreteness imagine that these two operators are the images of $R(r,s)\alpha\beta$ and $T(t,u)\gamma\delta$
under the bijection described in section  \ref{NPCorr}.
These two operators disagree in the placement of $J\sim O(\sqrt{N})$ boxes, since the excitation which has $J$ boxes
is located at corner $i$ for state $|\phi\rangle$ and at corner $j$ for state $|\psi\rangle$.
Thus, these two operators will start to mix at the $J$ loop order.
Further, for a non-zero intertwining map $I_{R^{(J)},T^{(J)}}$ we need to drop the boxes 
that disagree between the two operators\footnote{We use $R^{(J)}$ to denote a Young diagram obtained by dropping $J$
boxes from $R$ and similarly for $T^{(J)}$.}.
This implies that, after expressing permutations that appear in the dilatation operator in terms of adjacent swaps, only the
first term on the right hand side of (\ref{defrep}) contributes.
We need to swap all of the distant and local boxes which leads to a suppression of $O(N^{-1})$, for every
box in the excitation.
Consequently this mixing is completely suppressed at large $N$ and (\ref{nomix}) holds.

We are now ready to tackle (\ref{bigMix}).
Consider a localized excitation, located at corner $i$.
We study the mixing of this localized excitation with a delocalized excitation, that has $k$ boxes at corner $j\ne i$
and is otherwise located at corner $i$. 
These two operators disagree in the placement of at least $k$ boxes and so the first time they can possibly mix is at $k$
loops. 
To get a non-zero answer, for the interwining map, we need to drop the boxes that don't agree and this means that
we must keep terms in which distant boxes remain distant.
This again amounts to retaining the first term on the right hand side of (\ref{defrep}) and hence a suppression of 
$O(N^{-1})$, for every distant box.
Consequently this mixing is suppressed as $\sim N^{-k}$ at large $N$.
This demonstrates that (\ref{bigMix}) holds at large $N$.

We will end this section with a simple example illustrating the above argument.
The operators which mix are labeled by the Young diagrams shown in (\ref{opsmixing}).
They have a total of 2 $Y$ fields and many $Z$ fields.
\bea
A={\small \young(\,\,\,\,\,\,\,\,\,*,\,\,\,\,\,\,\,\,*Y,\,\,\,\,\,\,\,\,,\,\,\,\,\,\,\,\,,\,\,\,\,\,\,\,\,,\,\,\,\,\,\,\,\,,\,\,\,\,\,\,\,\,,\,\,\,\,\,\,\,*,Y)}\qquad
B={\small \young(\,\,\,\,\,\,\,\,\,*Y,\,\,\,\,\,\,\,\,*Y,\,\,\,\,\,\,\,\,,\,\,\,\,\,\,\,\,,\,\,\,\,\,\,\,\,,\,\,\,\,\,\,\,\,,\,\,\,\,\,\,\,\,,\,\,\,\,\,\,\,*)}\label{opsmixing}
\eea
$A$ and $B$ are the Young diagrams for the excitations in the background. They both have $O(N^2)$ boxes. 
The boxes labeled with a $Y$ correspond to $Y$ fields and they may be labeled 1. The boxes with a $*$ are $Z$ fields and may be labeled with $m+1=3$. For a non-zero answer, the states in $A$ which contribute have the bottom $Y$ labeled with a 1. Only in this case can we match the shape of $B$, after one box - the upper $Y$ box - is dropped. Using Dirac notation, the structure of the terms contributing to (\ref{Trtodo}) are
\bea
\sum_{i,j}\langle A',i|P_{+R,(+r,s)\mu_1\mu_2}(1,m+1)|A',j\rangle
\langle B',j|P_{+T,(+t,u)\nu_1\nu_2}(1,m+1)|B',i\rangle
\eea
It is clear that the only way that $\langle A',i|P_{+R,(+r,s)\mu_1\mu_2}(1,m+1)|A',j\rangle$ can be non-zero is if we keep the first term in (\ref{defrep}) when the permutation acts. Since the only boxes labeled with $m+1$ (the starred boxes) are distant from the bottom $Y$ box in $A$, this is suppressed as ${1\over N}$.

\section{Strong Coupling CFT}

In this section we want to explore the string theory interpretation of our results, adding to the discussion of the introduction.
The excitations we have considered in the CFT are all dual to excitations of the D3-brane giant gravitons that condensed to produce the geometry.
These are all open string excitations and we have demonstrated that they lead to emergent gauge theories.
In this section we will motivate why adding boxes to the Young diagrams give excitations that are localized to the brane, that is, why they are open strings.
There are also closed string excitations in the dual string theory. We will give an example of a closed string excitation.
For relevant earlier literature see \cite{Chen:2007gh,deMelloKoch:2009zm,Lin:2010sba,Diaz:2015tda}.

Why does adding extra boxes to a Young diagram as we have done above, lead to open strings excitations?
We can also phrase this question as: Why does adding extra boxes to a Young diagram lead to excitations localized on the branes?
Recall that there is an intimate connection between the entanglement of the underlying degrees of freedom and the geometry of spacetime. 
This is manifested in the Ryu-Takayanagi formula for entanglement entropy in terms of the area of a minimal 
surface\cite{Ryu:2006bv}. 
Further, Van Raamsdonk has conjectured that the amount of entanglement between two regions is related to the distance between them: the more the entanglement the less the distance between the two regions\cite{VanRaamsdonk:2010pw}.
For a recent relevant discussion see \cite{Simon:2018laf}.
To apply this to our set up, recall that the Young diagram is an instruction for how an operator composed of many fields is to be constructed.
Each box corresponds to a distinct field and the indices of fields in the same row are to be symmetrized, while the indices of fields in the same column are to be antisymmetrized.
This will in the end produce a highly entangled state, with fields corresponding to boxes that are nearby on the Young diagram being more entangled than boxes that are more distant.
The Young diagram becomes a convenient way to visualize the entanglement so that boxes that are nearby on the Young diagram, are nearby in spacetime.
To make these comments more precise we would need a better understanding of entanglement for multi part quantum systems.

If this interpretation is correct, then to produce a closed string excitation (which is not localized on the brane), we should construct an operator whose indices are not symmetrized or antisymmetrized with indices of the fields making up the background.
An example of such an operator is given by $O_{\{k\}}=\Tr (Y^{k_1}X^{k_2}Y^{k_3}\cdots)$.
Since this is a closed string state, we expect that the mixing of this operator with the background will correspond to
closed string absorption by a brane.
Intuition from a single brane suggests that this is highly suppressed because $g_s\sim O(N^{-1})$ at large $N$.
However, we are dealing with $O(N)$ branes so that we can't neglect mixing of $O_{\{k\}}$ with the background.
If this mixing were suppressed, we would be dealing with an $SU(2)$ sector of the planar Yang-Mills theory which is integrable.
We will explore this issue at strong coupling using string theory.

The state dual to $O_{\{k\}}$ should be a closed string moving in an LLM geometry. 
The general LLM geometry is described by the metric\cite{Lin:2004nb} $(i,j=1,2)$
\bea
ds^2 = -y(e^G+e^{-G})(dt + V_i dx^i)^2 + 
{1\over y(e^G+e^{-G})}(dy^2 + dx^i dx^i) + y e^{G} d\Omega_3 + y e^{-G} d\tilde{\Omega}_3
\label{LLMgeo}
\eea
where
\bea
 z = \tilde z + {1\over 2}= \frac{1}{2} \tanh(G) \qquad
y\partial_y V_i = \epsilon_{ij} \partial_j \tilde z\qquad
y(\partial_i V_j - \partial_j V_i) = \epsilon_{ij} \partial_y \tilde z
\eea
The metric is determined by the function $z$ which depends on the three coordinates $y, x^1$ and $x^2$ and is
obtained by solving Laplace's equation
\bea
\partial_i \partial_i z + y \partial_y \frac{\partial_y z}{y} = 0.  \label{LE}
\eea
In what follows we often trade $x^1,x^2$ for a radius and an angle, $r$ and $\varphi$.
Our focus is on geometries given by concentric black annuli on the bubbling plane.
For a set of rings with a total of $E$ edges with radii $R_l$ $l=1,2,...,E$ the geometry is determined by the 
functions\cite{Lin:2004nb}
\bea 
\tilde z =
\sum_{l=1}^E {(-1)^{E-l}\over 2}\left( {r^2+y^2-R_l^2\over\sqrt{(r^2+y^2+R_l^2)^2-4r^2 R_l^2}}-1\right),
\eea
\bea 
V_{\varphi}(x^1,x^2,y)=\sum_{l=1}^E {(-1)^{E-l+1}\over 2}\left( {r^2+y^2+R_l^2\over\sqrt{(r^2+y^2+R_l^2)^2-4r^2 R_l^2}}-1\right).
\eea
We need the $y=0$ limit of the metric, which is given by
\bea
   ds^2=-{1\over b}(dt+V_\varphi d\varphi)^2 +b (dy^2 +y^2 d\tilde{\Omega}_3^2)+b (dr^2+r^2d\varphi^2)+{1\over b} 
\left( \sin^2\psi d\beta^2+d\psi^2+\cos^2\psi d\alpha^2 \right)\cr
&&
\eea
with
\bea
  b(r)=\sqrt{\sum_{l=1}^E (-1)^{E-l}{R_l^2\over (R_l^2-r^2)^2}}
\eea
We look for classical string solutions to the equations of motion following from the Nambu-Goto action
\bea
S_{NG}={\sqrt{\lambda}\over 2\pi}\int d\tau L_{NG}
={\sqrt{\lambda}\over 2\pi}\int d\sigma \int d\tau\sqrt{(\dot X\cdot X')^2-\dot X^2 X^{\prime 2}}
\eea
The ansatz
\bea
t =  \tau\qquad \psi =\psi(\tau,\sigma)\qquad \alpha=\alpha(\tau,\sigma)\qquad
y=0\qquad r=0
\eea
with $\tilde\theta,\tilde\varphi,\tilde\psi,\varphi,\beta$ constant leads to a solution.
After inserting this into the equations of motion, the resulting equations describe a string moving on
\bea
   ds^2= {1\over b(0)}\left(-dt^2 +d\psi^2+\cos^2\psi d\alpha^2 \right)
\eea
This is string theory on ${\mathbb R}\times S^2$ which is integrable.
The single magnon solution is given by $t=\tau$, $\alpha =\tau+\sigma$ and
\bea
   \cos\psi ={\cos\psi_0\over\cos\sigma}\qquad -\psi_0\le\sigma\le\psi_0
\eea
The energy of this solution is given by
\bea
   E&=&{\sqrt{\lambda}\over 2\pi}\int_{-\psi_0}^{\psi_0}d\sigma {\partial L_{NG}\over\partial \dot{t}}\cr
     &=&{\sqrt{\lambda}\over \pi}{1\over b(0)}\cos\psi_0
\eea
This is the energy of a single magnon with $N\to N_{\rm eff}$ where
\bea
   N_{\rm eff}={N_1 (M + N_1) (M + N_1 + N_2)\over M^2 + N_1^2 + M (2 N_1 + N_2)}
\eea
and $N_1+N_2=N$.
In writing this formula we specialized to a geometry with a central black disk of area $N_1$, a white ring
of area $M$ and a black ring of area $N_2$.
If we take $N_2=O(1)=M$ at large $N$ we find $N_{\rm eff}=N_1=N(1+O(N^{-1}))$.
This is exactly as expected since this boundary condition corresponds to exciting so few giant gravitons that backreaction can be neglected and we must recover the AdS$_5\times$S$^5$ result as we have done.
The above result shows that the closed string is exploring the geometry at $r=0$ in the bubbling plane.
This region simply can't be explored by adding boxes to any corner of the background Young diagram. 
The result depends in a nontrivial way on the details of the background, as we might expect for an excitation that
is not localized on a specific set of branes. This supports our argument that this is a closed string excitation.
For this closed string excitation once again the only change as compared to the planar limit is
the replacement $N\to N_{\rm eff}$. This is probably only a property of the strong coupling limit.
Indeed, in the free theory the correlator of the closed string excitation and the background factorizes
\bea
\langle\chi)B(Z)\chi_B(Z)^\dagger O_{\{k\}}O_{\{k\}}^\dagger\rangle
=\langle\chi)B(Z)\chi_B(Z)^\dagger\rangle\langle O_{\{k\}}O_{\{k\}}^\dagger\rangle
\eea
which is not consistent with a simple $N\to N_{\rm eff}$ replacement.

\section{Summary and Outlook} 

In this article we have considered excitations of LLM geometries.
The excitations are constructed by adding boxes (representing the excitation) to a Young diagram with $O(N^2)$ boxes (representing the LLM geometry).
Adding a box to a row of a Young diagram implies that the indices of the added operator will be symmetrized or
antisymmetrized with the indices of adjacent boxes, so that the fields associated to the boxes added are highly entangled
with the fields associated to adjacent boxes.
Two objects that are entangled are nearby in spacetime, so that we produce excitations that are localized to the brane
worldvolume.   
These excitations are open strings and hence give rise to an emergent gauge theory.
We have constructed a bijection between operators in the Hilbert space of planar ${\cal N}=4$ super Yang-Mills and
operators in the planar Hilbert space of the emergent gauge theory.
Free field correlators of operators that are in bijection are related in a very simple way.
This immediately implies that since three point functions of single trace operators are suppressed in the planar limit of the
original free CFT, they are also suppressed in the planar limit of the free emergent gauge theory.
Since OPE coefficients are read from three point functions, the OPE coefficients vanish in the planar limit of both
free ${\cal N}=4$ super Yang-Mills and the free emergent gauge theory.
We have conjectured that this continues to be the case when interactions are turned on.
By considering the weak coupling CFT we have also given arguments concluding that the anomalous dimensions match
the dimensions of an ${\cal N}=4$ super Yang-Mills with gauge group $U(N_{\rm eff})$, where $N_{\rm eff}$ is read
from the factor of the boxes associated to the excitations.
Since any CFT is determined by its OPE coefficients and spectrum of anomalous dimensions, this strongly suggests that the planar limit of the emergent gauge theories are planar ${\cal N}=4$ super Yang-Mills theories.

We have been careful to stress that the planar limit of the emergent gauge theory agrees with planar ${\cal N}=4$ super Yang-Mills. The stronger statement, that the emergent gauge theory is ${\cal N}=4$ super Yang-Mills is not true: there are important differences between the two theories that are only apparent when going beyond the planar limit.
The emergent gauge theory has gauge group $U(N_{\rm eff})$. If this gauge theory really is ${\cal N}$=4 super Yang-Mills theory we expect a stringy exclusion principle cutting off the angular momentum of the giant graviton at momentum $N_{\rm eff}$. In actual fact, the maximum angular momentum for a giant graviton is in general below this and it is set by the shape of the background Young diagram. Similarly, dual giant gravitons can usually have an arbitrarily large angular momentum. In the emergent gauge theory, the dual giant must fit inside the corner at which the emergent gauge theory is located, so there are no dual giant excitations with arbitrarily large momentum.
These discrepancies arise because the giant graviton excitations detect the structure of the bubbling plane.
They can probe the difference between a black disk in a sea of white or just one ring among many or something else.
So even in the large $N$ limit, the emergent gauge theory and ${\cal N}=4$ super Yang-Mills theory are different.
They do however share the same planar limit.

An interesting technical result that has been achieved is the description of states when some of the rows of the Young
diagram describing the giant graviton branes are equal in length.
Previous studies \cite{Carlson:2011hy,Koch:2011hb,deMelloKoch:2012ck} have considered the displaced corners approximation in which the length between any two rows (for a system of dual giant gravitons) scales as $N$ in the large $N$ limit. In this situation, the action of the symmetric group simplifies and explicit formulas for the restricted characters can be developed\cite{Carlson:2011hy,Koch:2011hb}.
Here we have the case that many row lengths are of comparable size. Progress is achieved by uncovering the relationship between the relevant restricted Schur computations and those of the planar limit.
We also allowed some $Z$ fields in the excitation which includes the case that the row lengths are similar but not identical.

There are a number of interesting directions that could be pursued.
First, perhaps there are new holographic dualities: each emergent gauge theory might itself be dual to an AdS$_5\times$S$^5$ geometry, in a suitable limit.
There maybe a limit of the geometry that zooms in on the edge of the black regions in the bubbling plane to give an 
AdS$_5\times$S$^5$ geometry with $N_{\rm eff}$ units of five form flux.
Restricting to excitations that belong to ${\cal H}^{(i)}_{\rm CFT}$ is how we restrict to the integrable subsector
in the CFT.
The limit that isolates an AdS$_5\times$S$^5$ geometry would restrics us to the integrable subsector in the string theory.
This is currently under active investigation \cite{MR}.

There are a number of questions we could pursue to further explore the dynamics of the emergent gauge theory.
As we have mentioned, the worldvolume of the giant gravitons is a distinct space from the space on which
the original CFT is defined.
How is locality in this emergent space of the emergent gauge theory realized? This may provide a simple testing ground
for ideas addressing the emergence of spacetime.
We have argued that there are integrable subsectors in large $N$ but non-planar limits of ${\cal N}=4$ super Yang-Mills.
Can we find further evidence for the integrability of these subsectors? Even more important, how can this integrability be exploited to explore the physics of emergent gauge theories in interesting and non-trivial ways?
For other promising indications of integrability beyond the planar limit see \cite{Eden:2017ozn,Bargheer:2017nne,Bargheer:2017nne}.
Besides the local observables we have considered, the emergent gauge theory will have Wilson loops.
It maybe interesting to explore these non-local observables.

The emergent gauge theory that we have explored in this article is only a decoupled sector at large $N$.
What are the first corrections which couple the emergent gauge theory to the rest of the theory? 
Presumably these corrections correspond to closed string absorption/emission by branes.
This is something concrete that can be evaluated.

Finally, decoupling limits for gauge theory living on the intersections of giant gravitons have been considered 
in \cite{Balasubramanian:2007bs,Fareghbal:2008ar,Harmark:2008gm,deBoer:2011zt,Johnstone:2013eg}.
It would be interesting to see if the methods developed in this article can be used to clarify the emergent gauge theories arising in these cases, which may shed light on the microstates of near-extremal black holes in AdS$_5\times$S$^5$.

{\vskip 0.5cm}
\noindent
\begin{centerline} 
{\bf Acknowledgements}
\end{centerline} 

This work is supported by the South African Research Chairs
Initiative of the Department of Science and Technology and National Research Foundation
as well as funds received from the National Institute for Theoretical Physics (NITheP).
J.-H.Huang is supported by the Natural Science Foundation of Guangdong Province (No.2016A030313444).
We are grateful for useful discussions to Shinji Hirano, Minkyoo Kim, Sanjaye Ramgoolam and Jaco Van Zyl.

\begin{appendix}

\section{Ratios of hooks}\label{ProveIdentity}

The hook $H_R(i,j)$ of the box in row $i$ and column $j$ of $R$ is the set of boxes $(a,b)$ with $a=i$ and 
$b\ge j$ or $a\ge i$ and $b=j$. The hook-length $h_R(i,j)$ is the number of boxes in the hook $H_R(i,j)$.
To visualize the hook associated to a given box, imagine an elbow with its joint in the box and one arm exiting $R$ by
moving to the right through the row of the box and one arm exiting by moving down through the column.
The hook length is the number of boxes the elbow passes through.
We use ${\rm hooks}_R$ to denote the product of hook lengths for each box in $R$.
In this Appendix we want to derive a formula for the ratio
\bea
{{\rm hooks}_{+R}\over {\rm hooks}_{+r}}
\eea
$+r$ is obtained from $+R$ by removing a total of $|R|-|r|$ boxes.
All of these boxes are located close to corner $i$ of Young diagram $B$.

Start by removing a single box from $+R$ to obtain the Young diagram $+R'$.
Consider the ratio
\bea
{{\rm hooks}_{+R}\over {\rm hooks}_{+R'}}
\eea
Imagine that the box that was removed comes from row $a$ and column $b$ of $R$.
Denote the length of row $a$ by $l_a$ and the length of column $b$ by $l_b$.
The numbers $a,b,l_a,l_b$ are all much smaller than $\sqrt{N}$.
Most hook lengths in the numerator will equal the hook lengths in the denominator. 
The only hook lengths that don't match are lengths for hooks that enter or exit through the box that is removed.
After many cancellations we find
\bea
{{\rm hooks}_{+R}\over {\rm hooks}_{+R'}}=
\prod_{j=1}^i {L(j,i)-b+l_B\over L(j,i)-N_j-b+l_b}
\prod_{l=i+1}^C {L(i+1,l)-a+l_a\over L(c+1,l)-M_l-a+l_a}{{\rm hooks}_R\over {\rm hooks}_{R'}}
\eea
where
\bea
  L(c,d)=\sum_{k=c}^d (N_k+M_k)
\eea
and $N_k$ and $M_k$ are defined in Figure \ref{LLM}.
These numbers specify the background Young diagram.
In the large $N$ limit this result can be simplified to
\bea
{{\rm hooks}_{+R}\over {\rm hooks}_{+R'}}=\eta_B
{{\rm hooks}_R\over {\rm hooks}_{R'}}\left( 1+O\left({1\over N}\right)\right)
\eea
where
\bea
   \eta_B=\prod_{j=1}^i {L(j,i)\over L(j,i)-N_j}\prod_{l=i+1}^C {L(i+1,l)\over L(c+1,l)-M_l}
\eea
Notice that $\eta_B$ is independent of $a$ and $b$, at large $N$.
If we have removed two boxes from $+R$ to obtain $+R''$, we can use the above result to compute
\bea
{{\rm hooks}_{+R}\over {\rm hooks}_{+R''}}&=&
{{\rm hooks}_{+R}\over {\rm hooks}_{+R'}}{{\rm hooks}_{+R'}\over {\rm hooks}_{+R''}}\cr
&=&\left(\eta_B\, {{\rm hooks}_{R}\over {\rm hooks}_{R'}}\right)
\left(\eta_B {{\rm hooks}_{R'}\over {\rm hooks}_{R''}}\right)\left( 1+O\left({1\over N}\right)\right)\cr
&=&(\eta_B)^2{{\rm hooks}_{R}\over {\rm hooks}_{R''}}\left( 1+O\left({1\over N}\right)\right)
\eea
At large $N$, every time we remove a box from $+R$ it results in a factor of $\eta_B$ in the ratio
of hooks lengths.
We have to remove $|R|-|r|$ boxes from $R$ to obtain $r$, so that we find
\bea
{{\rm hooks}_{+R}\over {\rm hooks}_{+r}}={{\rm hooks}_{R}\over {\rm hooks}_{r}}
(\eta_B)^{|R|-|r|}\left( 1+O\left({1\over N}\right)\right)
\eea
which is the identity (\ref{HIdentity}) used in section \ref{NPCorr}.

\section{Ratios of factors}\label{Ratf}

Recall that $f_R$ denotes the product of the factors of each box in $R$ and that a box in row $i$ and column $j$
has factor $N-i+j$.
In this Appendix we will compute the ratio of the product of factors for a Young diagram $+R$ and Young diagram $B$.
$+R$ is obtained by attaching a smaller Young diagram $R$ to the Young diagram $B$.
The argument is rather simple and most easily illustrated with an explicit example.
\begin{center}
\begin{figure}[!h]
\centering
\includegraphics[width=0.6\textwidth]{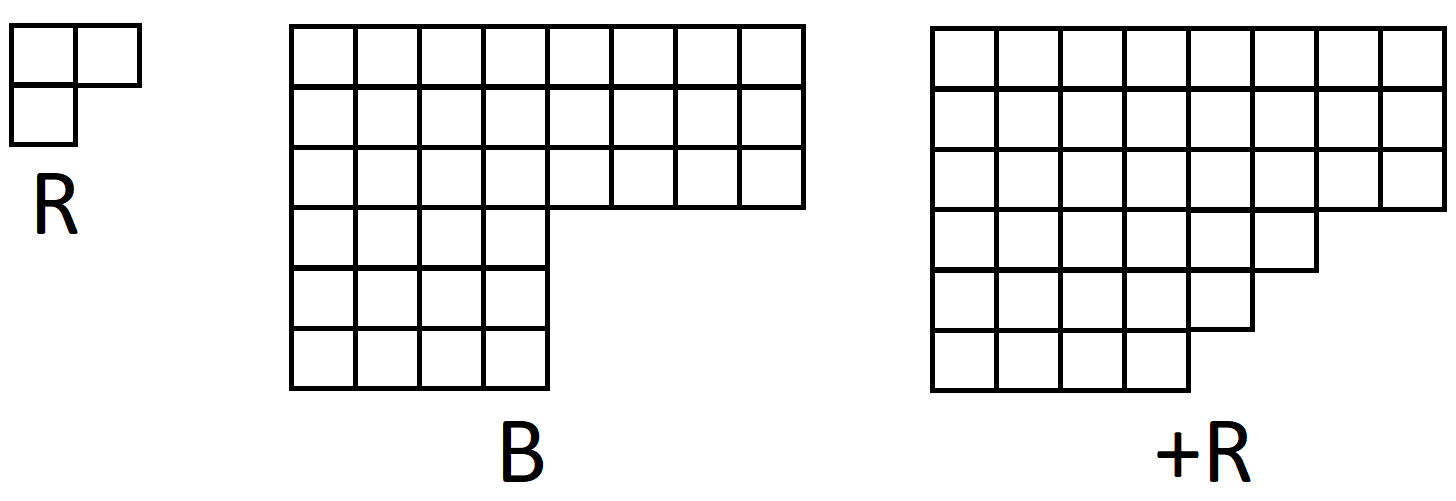}
\caption{An example showing Young diagrams $R$, $B$ and $+R$. The Young diagram $+R$ is obtained
by stacking $R$ next to $B$.}
\label{ratios}
\end{figure}
\end{center}

Consider the Young diagrams shown in Figure \ref{ratios} above.
It is simple to see that
\bea
   f_R=N(N-1)(N+1)
\eea
and
\bea
   {f_{+R}\over f_B}=(N+\delta)(N+\delta-1)(N+\delta+1)
\eea
where $\delta=1=5-4$.
In general, if the top most and left most box of $R$ is added to row $a$ and column $b$ of $B$, we will have
$\delta = b-a$.

\section{Delocalized Trace Structures are Preserved}\label{deloc}

In this Appendix we compute correlation functions of delocalized operators.
Our results suggest that, in general, there is no simple relationship between correlation functions of delocalized 
operators and correlation functions of operators in the planar limit, even in the free CFT.
The results of our computation do however provide evidence that mixing between different 
trace structures is suppressed, even for the delocalized operators.

To keep the discussion simple consider operators constructed from a single field $Z$.
This will already probe aspects of the operator mixing issue.
As a simple warm up example, consider delocalized excitations constructed by starting with operators of the form 
${\rm Tr}(\sigma_1 Z^{\otimes n_1}){\rm Tr}(\sigma_2 Z^{\otimes n_2})$.
To construct a delocalized excitation, begin by writing
\bea
{\rm Tr}(\sigma_1 Z^{\otimes n_1})=\sum_{R_1\vdash n_1}\chi_{R_1}(\sigma_1)
\chi_{R_1}(Z)\cr\cr
{\rm Tr}(\sigma_2 Z^{\otimes n_2})=\sum_{R_2\vdash n_2}\chi_{R_2}(\sigma_2)
\chi_{R_2}(Z)
\eea
The delocalized excitation is given by
\bea
O^{(B)}(\sigma_1,\sigma_2)=\sum_{R_1,R_2}\chi_{R_1}(\sigma_1)\chi_{R_2}(\sigma_2)\chi_{+(R_1,R_2)}(Z)
\eea
The Young diagram $+(R_1,R_2)$ is obtained by adding $R_1$ at the $i$th inward pointing corner and adding
$R_2$ at the $j$th inward pointing corner.
This corresponds to localizing ${\rm Tr}(\sigma_1 Z^{\otimes n_1})$ at the $i$th corner and localizing 
${\rm Tr}(\sigma_2 Z^{\otimes n_2})$ at the $j$th corner.
It is now rather simple to evaluate the correlator
\bea
&&\langle O^{(B)}(\sigma_1,\sigma_2)(x_1)O^{(B)}(\tau_1,\tau_2)^\dagger(x_2)\rangle_B\cr\cr
&&=\sum_{R_1\vdash n_1,R_2\vdash n_2}\chi_{R_1}(\sigma_1)\chi_{R_2}(\sigma_2)
\chi_{R_1}(\tau_1)\chi_{R_2}(\tau_2) {f_{+(R_1,R_2)}\over f_B |x_1-x_2|^{2n_1+2n_2}}\cr\cr
&&=\sum_{R_1\vdash n_1}\chi_{R_1}(\sigma_1)\chi_{R_1}(\tau_1){f_{R_1}(N_{\rm eff,1})\over |x_1-x_2|^{2n_1}}
\sum_{R_2\vdash n_2}\chi_{R_2}(\sigma_2)\chi_{R_2}(\tau_2){f_{R_2}(N_{\rm eff,2})\over |x_1-x_2|^{2n_2}}\cr\cr
&&=\langle {\rm Tr}(\sigma_1 Z^{\otimes n_1})(x_1){\rm Tr}(\tau_1 Z^{\otimes n_1})(x_2)\rangle_{N\to N_{\rm eff,1}}
\langle {\rm Tr}(\sigma_2 Z^{\otimes n_2})(x_1){\rm Tr}(\tau_2 Z^{\otimes n_2})(x_2)\rangle_{N\to N_{\rm eff,2}}\cr
&&\label{delocres}\eea
In the above expression, $f_R(M)$ means the product of the factors of Young diagram $R$ with $N$ replaced by $M$.
Further $N_{\rm eff,1}$ is the factor of the first box added to corner $i$ and $N_{\rm eff,2}$ is the factor of the first 
box added to corner $j$.
The above result implies that the delocalized correlator has factorized into two factors, one for each corner on which
the operator is located.
Each factor is a correlation function.
The value of $N$ is replaced by an effective value of $N$ for each corner.
It is worth emphasizing that the expressions on the last line of (\ref{delocres}) are exact.
This result implies that trace mixing is even more constrained for the delocalized excitation than it is in the planar limit.
Indeed, in the planar limit we will have mixing if the trace structure of 
${\rm Tr}(\sigma_1 Z^{\otimes n_1}){\rm Tr}(\sigma_2 Z^{\otimes n_2})$ matches the trace
structure of ${\rm Tr}(\tau_1 Z^{\otimes n_1}){\rm Tr}(\tau_2 Z^{\otimes n_2})$.
For the delocalized excitation we will only have mixing if the trace structure of ${\rm Tr}(\sigma_1 Z^{\otimes n_1})$
matches ${\rm Tr}(\tau_1 Z^{\otimes n_1})$ and the trace structure of ${\rm Tr}(\sigma_2 Z^{\otimes n_2})$ matches
${\rm Tr}(\tau_2 Z^{\otimes n_2})$.

There is a second type of delocalized excitation we could consider: a single trace operator that is itself delocalized.
As an example, consider a single trace operator that is distributed between corners $i$ and $j$.
To write such a loop we introduce the space time independent auxiliary field ${\cal X}^a_b$, which has two point
function
\bea
\langle {\cal X}^a_b {\cal X}^c_d\rangle = \delta^a_d\delta^c_b
\eea
Using this auxiliary field we can split any single trace operator into two traces, that reassemble to give a single
trace when the average over ${\cal X}$ is performed.
For example, we can replace
\bea
   {\rm Tr}(Y^5)\quad\longrightarrow\quad{\rm Tr}(Y^2 {\cal X}){\rm Tr}(Y^3 {\cal X})\label{SpOp}
\eea
Performing the average over ${\cal X}$, we recover our original loop
\bea
  \langle {\rm Tr}(Y^2 {\cal X}){\rm Tr}(Y^3 {\cal X})\rangle
=(Y^2)^b_a\,\, (Y^3)^d_c \langle {\cal X}^a_b {\cal X}^c_d\rangle
={\rm Tr}(Y^5)
\eea
The advantage of splitting things in this way, is that we can now follow exactly the same logic that we used for the first
example above.
We will take this to be the definition of the delocalized single trace operator.
For the general operator constructed from $Y$s, the resulting expression is of the form\footnote{Imagine that our
operator is constructed using $n$ $Z$s. The restricted Schur polynomials needed for this computation involve restricting 
$S_{n+1}$ to $S_n$. There is no need for multiplicity labels when studying this restriction.}
\bea
O_A(Y)=\sum_{R^1,R^2,r^1,r^2}
a^{(A)}_{R^1,R^2,r^1,r^2}
\chi_{R^1,(r^1,{\tiny\yng(1)})}(Y,{\cal X})
\chi_{R^2,(r^2,{\tiny\yng(1)})}(Y,{\cal X})
\eea
The single extra box in the labels for the restricted Schur polynomial represents the auxiliary ${\cal X}$ field.
We can now, following the example we studied above, attach $R_1$ and $R_2$ to different corners and in this 
way obtain the delocalized single trace operator.
For operators that involve more than two corners, we would need to introduce more than one auxiliary field.
Concretely, for the case we consider, we have
\bea
O^{(B)}_A&=&\sum_{R^1,R^2,r^1,r^2}\,\,
a^{(A)}_{R^1,R^2,r^1,r^2}\,\,
\chi_{+(R^1,R^2),\left(+(r^1_1,r^1_2),{\tiny\yng(1)\times\yng(1)}\right)}(Z,Y,{\cal X})\label{dlo}
\eea
The notation ${\tiny \yng(1)}\times{\tiny \yng(1)}$ is just to reflect the fact that we have not organized the auxiliary fields into
representations of $S_2$.
We can now average over ${\cal X}$ in (\ref{dlo}) to obtain an operator that does not depend on the auxiliary fields.
This averaging is easily performed using the methods developed in \cite{Koch:2008cm}.
It is straight forward, but tedious and messy, to check that mixing between different trace structures of these delocalized
excitations is also suppressed.

Lets illustrate the above construction with the simplest possible example: we consider two delocalized operators.
The first, $O_A$, is given by placing ${\rm Tr}(Y)$ at corner $i$ and ${\rm Tr}(Y)$ at corner $j$.
The second, $O_B$, is obtained by distributing ${\rm Tr}(Y^2)$ between the two corners.
When the background is not present, the relevant correlators are
\bea
\langle {\rm Tr}(Y^2)(x_1) {\rm Tr}(Y^{\dagger 2})(x_2)\rangle &=& {2N^2\over |x_1-x_2|^4}\cr
\langle {\rm Tr}(Y)^2(x_1) {\rm Tr}(Y^\dagger)^2(x_2)\rangle &=& {2N^2\over |x_1-x_2|^4}\cr
\langle {\rm Tr}(Y)^2(x_1) {\rm Tr}(Y^{\dagger 2})(x_2)\rangle &=& {2N\over |x_1-x_2|^4}
\eea
It is clear that the last correlator, which mixes different trace structures, is down by a factor of $N$. 
If we had normalized the two point functions to one, the last correlator above vanishes at large $N$ which 
shows that different trace structures don't mix. 
The delocalized operator with ${\rm Tr}(Y)$ at corner $i$ and ${\rm Tr}(Y)$ at corner $j$ is obtained by adding a
single box at corner $i$ of $B$ and a single box at corner $j$.
Denote the factor of the box added at corner $i$ by $N_{\rm eff,1}$ and the factor of the box added at corner $j$ by 
$N_{\rm eff,2}$.
It is a simple matter to find
\bea
\langle O_A^{(B)}(x_1) O_A^{(B)}(x_2)^\dagger\rangle
=\eta_B\tilde\eta_B {N_{\rm eff,1}N_{\rm eff,2}\over |x_1-x_2|^4}
\eea
in complete agreement with (\ref{delocres}).
The coefficient $\eta_B\tilde\eta_B$ is an order 1 number that arises from computing the ratios of hooks.
After averaging over the ${\cal X}$ fields we find that $O_B^{(B)}(x_2)$ is a sum of two terms.
One is clearly leading and has coefficient $\sqrt{1-{1\over (N_{\rm eff,1}-N_{\rm eff,2})^2}}$.
The subleading term have coefficient ${1\over N_{\rm eff,1}-N_{\rm eff,2}}$.
The leading term involves a twisted character in the notation of \cite{deMelloKoch:2007uu}, while the subleading
term is a normal restricted character.
We find that both terms contribute to the correlator
\bea
\langle O_B^{(B)}(x_1) O_B^{(B)}(x_2)^\dagger\rangle
=\eta_B\tilde\eta_B{N_{\rm eff,1}N_{\rm eff,2}\over |x_1-x_2|^4}
\eea
while only the subleading term contributes to the mixed correlator
\bea
\langle O_A^{(B)}(x_1) O_B^{(B)}(x_2)^\dagger\rangle
=\eta_B\tilde\eta_B{N_{\rm eff,1}N_{\rm eff,2}\over (N_{\rm eff,1}-N_{\rm eff,2})|x_1-x_2|^4}
\eea
Since $N_{\rm eff,1}-N_{\rm eff,2}$ is of order $N$, this clearly demonstrates the suppression.
Although there is little doubt that mixing between different trace structures is suppressed for the general delocalized 
excitations, at this point in time we do not have a simple general argument for this conclusion.

\section{Localized and Delocalized Mixing at One Loop}

In this Appendix we study a simple example of mixing between a localized and a delocalzied operator at one loop.
Since we don't want a selection rule to prevent the operators from mixing, we need to consider operators 
that differ in the placement of at most one box.
To make the computation as transparent as possible choose particularly simple operators.
Our goal is to show that this mixing is of order $N^{-1}$.
This is a simple illustration that the mixing between a delocalized operator and a local operator, 
is suppressed at large $N$.

The local operator that we consider is $O_{+{\tiny\yng(2,2)},(+{\tiny\yng(2)},{\tiny \yng(2)})}(Z,Y)$. 
The representation ${\tiny\yng(2,2)}$ produces $(+{\tiny\yng(2)},{\tiny \yng(2)})$ once upon restricting from
$S_4$ to $S_2\times S_2$ so that there is no need for multiplicity labels.
Lets assume that this excitation is localized at corner $i$.
For the delocalized operator, we assume that we have $({\tiny\yng(2,1)},({\tiny\yng(2)},{\tiny \yng(1)}))$
at corner $i$ and ${\tiny \yng(1)},(\cdot,{\tiny \yng(1)})$ at corner $j$.
For this example we can evaluate the matrix element (\ref{matelem}) exactly.
The result is
\bea
N_{+{\tiny\yng(2,2)},(+{\tiny\yng(2)},{\tiny\yng(2)});
+\left({\tiny\yng(2,1)}_i,{\tiny \yng(1)}_j\right),(+{\tiny\yng(2)}_i,\left({\tiny\yng(1)}_i,{\tiny\yng(1)}_j)\right)}
={\lambda_{\rm eff,i}\over 4\pi}\sqrt{N_{\rm eff,i}\over N_{\rm eff,j}}
{1\over N_{\rm eff,j}-N_{\rm eff,i}}\left(1+O\left({1\over N}\right)\right)
\eea
where $N_{\rm eff,i}$ is the factor of the first box added at corner $i$, $N_{\rm eff,j}$ is the factor of the first box 
added at corner $j$ and $\lambda_{\rm eff,i}\equiv g_{YM}^2 N_{\rm eff,i}$.
The fact that this mixing is of order $N^{-1}$ is in perfect accord with the arguments of section \ref{NBMix}.

\section{Correcting the planar limit}

The emergent gauge theory has 't Hooft coupling $g_{YM}^2 N_{\rm eff}$ with $g_{YM}^2$ the coupling of the original
CFT. It is natural to ask if (non-planar) higher genus corrections are suppressed by powers of $N$ or powers of $N_{\rm eff}$. This Appendix gives a discussion of the issue. 

The article \cite{deMelloKoch:2009jc} studied excitation of the annulus LLM background, with boundary condition given by a single black annulus (of area $N$) with a central white disk (of area $M$). The Young diagram describing this geometry has a total of $N$ rows and $M$ columns.
A simple and clean argument shows that the 1/2 BPS correlators, with excitations constructed using only $Z$ fields, admit an expansion with $N_{\rm eff}^{-2}$ playing the role of the genus counting parameter\cite{deMelloKoch:2009jc}.
In the 1/2 BPS sector, this result generalizes to multi ring geometries and again the genus counting parameter is 
$N_{\rm eff}^{-2}$.

To go beyond the half BPS sector the result (\ref{CoolR}) can be used. After rescaling the fields which are not $Z$ fields, by a factor of $1/\sqrt{\eta_B}$, we find a product of two terms
\bea
\langle O_A (x_1)O_B(x_2)^\dagger\rangle_B &=& F_{AB}(N_{\rm eff}){1\over |x_1-x_2|^{2J}}
\left( 1+O\left({1\over N}\right)\right)
\eea
The first factor on the RHS above admits an expansion in $N_{\rm eff}^{-1}$.
The second factor does not. Thus, in general our amplitude can't be developed as a series in the two small parameters
$\lambda_{\rm eff}$ and $N_{\rm eff}^{-2}$.

\end{appendix}

{} 
\end{document}